\documentclass[a4paper,12pt]{article}
\usepackage[utf8]{inputenc}
\usepackage{cancel}
\usepackage{ulem}
\usepackage{amsfonts}
\usepackage{amssymb}
\usepackage{graphicx}
\usepackage{amsmath}
\usepackage{enumerate}
\usepackage{here}
\usepackage{subfloat}
\usepackage{subfig}
\usepackage{color}
\usepackage{float}
\usepackage{here}
\usepackage{cite}
\usepackage{mathrsfs}
\usepackage{float,epsfig}
\usepackage{multirow}
\usepackage{dcolumn}
\usepackage{placeins}
\usepackage{graphicx}
\usepackage{bm}
\usepackage{amsmath,amssymb,amsthm}
\usepackage[colorlinks=true,linkcolor=blue]{hyperref}
\usepackage[table]{xcolor}
\title{\bf \Large  Thermodynamic criticality of d-dimensional charged AdS black holes surrounded by quintessence with a cloud of strings background}

\author{M. Chabab$^{1}$\footnote{mchabab@uca.ac.ma (Corresponding author)}, S. Iraoui$^{1}$\footnote{samir.iraoui@ced.uca.ma}\\
	\\ 
	{\small $^{1}$ High Energy and Astrophysics Laboratory, Physics Department, FSSM, Cadi Ayyad University, 
	}\\
	{\small  P.O.B. 2390 Marrakech, Morocco.
	}
}
\date{}
\begin{document} \maketitle

\begin{abstract}
We focus on the study of the exact solution corresponding to charged AdS black holes surrounded by quintessence with a cloud of strings present in higher dimensional spacetime. We then investigate its corresponding thermodynamic criticality  in the extended phase space and  show that the spacetime  dimension has no effect on the existence of small/large phase transition for such black holes. The heat capacity is evaluated and the geothermodynamics of Quevedo analyzed for different spacetime dimensions with the cloud of strings and quintessence parameters. We calculate the critical exponents describing the behavior of relevant thermodynamic quantities near the critical point.  Finally, we  also discuss the uncharged case, show how it is sensitive to the quintessence and  strings cloud parameters, and when the thermodynamic behavior of the uncharged black holes is similar to Van der Waals fluid.
\end{abstract}

{  Keywords:    \small   Black holes;  Phase transitions; Quintessence; Cloud of strings.     }

  \section{Introduction}
  
  High precision astronomical observations have shown that the universe is undergoing a phase of accelerated expansion \cite{Riess,Perlmutter}, which might be due to dark energy acting as a repulsive gravity. A possible origin of this phenomena comes from the so-called quintessence,  with a state equation formulated through the relation between the negative pressure and the  energy density as $p=\omega_q \rho_{q}$ \cite{Caldwell}. The first general solution of spherically symmetric static Einstein equations in four dimensions for the quintessence satisfying the condition of the additivity and linearity was constructed by Kiselev \cite{Kiselev}. Then, this solution was extended  to $d$-dimensional spacetime by assuming that quintessence lives not only on the brane, but  in the bulk \cite{Chen}. Later, other extensions of the Kiselev solutions were obtained and studied, like the case of charged black holes \cite{Azreg}, $5d$ black hole in Einstein Gauss-Bonnet gravity  \cite{Ghosh5D}, $d$ dimensional Lovelock gravity \cite{GhoshLovlock}.
  
  The recent theoretical developments are in favor of a scenario which represent universe as a collection of extended object instead of point-like particles. The most natural and popular candidate is a one-dimensional continua strings object. We emphasize that the study of Einstein's equations coupled with a cloud of strings may be very important since relativistic strings at a classical level can be used to construct applicable models. The general solution for a cloud of strings with spherical symmetry was first studied by Letelier \cite{Letelier}. Then, its generalization for third-order Lovelock gravity  \cite{Ghosh2014} and  the solution for Einstein-Gauss-Bonnet theory in the Letelier spacetime  \cite{Herscovich} were formulated. Afterwards, studies combining  the dark matter in the quintessence form with the cloud of strings were recently proposed for charged AdS black hole  \cite{Toledo2019} as well as Lovelock gravity \cite{Toledo2018Lovelok}. 
  
  Lately, the investigations of black holes phase transitions in asymptotically Anti de-Sitter spacetime have aroused growing interest, since these transitions have been related to holographic superconductivity in the context of the AdS/CFT correspondence \cite{Maldacena}. The studies of these black holes criticality have generally identified the cosmological constant with thermodynamic pressure and its conjugate with thermodynamic volume \cite{Belhaj,Belhaj2,Chabab,Kubiznak}. In addition, other endeavors have been made to establish relationship between the thermodynamic phase transitions and the particle behavior in the curved spacetime around the black holes \cite{ChababQNM,ChababQNM2,Chababphot,Liu}.  To investigate the critical behavior of black holes, two methods are usually called upon:  The analysis of the heat capacity and the geothermodynamics approaches proposed by Ruppeiner \cite{hep49}, Quevedo \cite{Quevedo} or  HPEM \cite{Newmetric}. The phase transitions are generally revealed when discontinuities show up in a state space variable.

  This paper is organized as follows. In the next section, we present a generalized  solution  corresponding to charged AdS black holes surrounded by quintessence with a cloud of strings in higher dimensional spacetime. In section \ref{sec3},  we discuss its thermodynamics and  briefly summarize its critical behavior. In section \ref{sec4}, we use the geothermodynamics approach of Quevedo to investigate criticality of $d$-dimensional RN-AdS black hole surrounded by quintessence with cloud of strings. In section \ref{sec5}, we calculate the critical exponents and show that they coincide with those of Van der Waals system. We devote section \ref{sec6} to consider the particular case of uncharged black hole in these backgrounds and  its thermal behavior sensitivity to quintessence and strings cloud parameters. The last section is devoted to our conclusion.
  
  \section{Quintessence surrounding d-dimensional RN-AdS black holes with a cloud of strings}\label{sec2}
  First, we consider  the metric for AdS asymptotically space-time in $d$ dimension  generated by a charged static black hole  and surrounded by a cloud of strings and quintessence. The energy-momentum tensor of the cloud of strings in an arbitrary dimension can be written as, \cite{Letelier,Ghosh2014}
  \begin{equation}\label{TmnString}
  T_{\mu}^{\ \nu}=\frac{a}{r^{d-2}}Diag\left[1,1,0,......,0\right],
  \end{equation}
  while that corresponding to the quintessence dark energy reads, \cite{Kiselev,Chen},
  \begin{equation}\label{TmnQuintrt}
  T_{t}^{\ t}=T_{t}^{\ t}=\rho_{q},
  \end{equation}
  \begin{equation}\label{TmnQuinttheta}
  T_{\theta_{i}}^{\theta_{i}}=-\frac{\rho_{q}}{d-2}\left((d-1)\omega_{q}+1\right), \ i=(1,...,d-2),
  \end{equation}
  Here $a$ is an integration constant related to the presence of the cloud of strings, $\rho_{q}$ the energy density for quintessence and $\omega_{q}$ represents the quintessential state parameter. Taking into account the presence of an electromagnetic field, we have:
  \begin{equation}\label{Tmncharge}
  T_{\mu\nu}=F_{\mu\sigma}F_{\nu}^{\sigma}-\frac{1}{4}g_{\mu\nu}F_{\lambda\sigma}F^{\lambda\sigma},
  \end{equation}
  where $F_{\mu\nu}=\partial_{\mu}A_{\nu}-\partial_{\nu}A_{\mu}$ is the electromagnetic tensor and $A_{\mu}$  the electric gauge potential.
  
  Following \cite{Toledo2018Lovelok}, we assume that the quintessence and cloud of strings do not interact with each other. Therefore, the resulting energy-momentum tensor must be a linear superposition.  By assuming  the following ansatz for the metric,
  \begin{equation}\label{lineelement}
  ds_{d}^{2}=-f(r)dt^{2}+f(r)^{-1}dr^{2}+r^{2}d\Omega^{2}_{d-2},
  \end{equation} 
  where $d\Omega^{2}_{d-2}$ denotes the metric on unit $\left(d-2\right)$-sphere, one can eliminate $\rho_{q}$ from the Einstein field equations, $G_{\mu}^{\ \nu}+\Lambda g_{\mu}^{\ \nu}= \sum T_{\mu}^{\ \nu}$,  and get the following differential equation,

  \begin{equation}\label{differential}
  r^2 f''(r)+\mathcal{F}_1 r f'(r)+\mathcal{F}_2 \left(f(r)-1\right) +\mathcal{F}_3  r^2+\mathcal{F}_4 r^{-2\left(d-3\right)}+\mathcal{F}_5 r^{-\left(d-4\right)}=0,
  \end{equation} 
  with
  \begin{align*}\label{coeff}
  \mathcal{F}_1& =\left(\left( d-1\right) \omega_q +2 d-5\right), \\
  \mathcal{F}_2& =\left(d-3\right) \left(\left(d-1\right) \omega_q +d-3\right),\\
  \mathcal{F}_3& =\Lambda \frac{2 \left(d-1\right)  \left(\omega_q +1\right)}{d-2}, \\
  \mathcal{F}_4& =q^2 \left(d-3\right) \left(\left(d-1\right) \omega_q -d+3\right),\\
  \mathcal{F}_5& =\frac{2 \left(\left(d-1\right) \omega_q +1\right) a}{d-2}.
  \end{align*} 
  Note that the cosmological constant $\Lambda$  is negative as required by the $AdS$ space.  Here the Maxwell equations, $\nabla_{\nu} \left(\sqrt{-g}F^{\mu\nu}\right)=0$ are used to evaluate the electric potential.
  
  The solution for such equations is given by,
  \begin{equation}\label{exactsolution}
  f(r)=1-\frac{m}{r^{d-3}}+\frac{q^{2}}{r^{2\left(d-3\right)}}-\frac{2 \Lambda}{\left(d-2\right)\left(d-1\right)}r^{2}-\frac{\alpha}{r^{\left(d-1\right)\omega_{q}+d-3}}-\frac{2 a}{\left(d-2\right)r^{d-4}},
  \end{equation} 
  where $\alpha$ is a  positive normalization factor, related to the density of quintessence, through the expression:
  \begin{equation}\label{factorQuint}
  \rho_{q}=-\frac{\alpha \omega_{q}\left(d-1\right)\left(d-2\right)}{4 r^{\left(d-1\right)\left(\omega_{q}+1\right)}},
  \end{equation} 
  $m$ and $q$ are integration constants proportional to the mass  and  charge of the black hole respectively given by the following formulas: \cite{Chamblin,Mann}:
  \begin{equation}\label{ADMmass}
  M=\frac{d-2}{16 \pi}\Omega_{d-2} m,\ Q=\frac{\sqrt{2\left(d-2\right)\left(d-3\right)}\Omega_{d-2}}{8 \pi}q,
  \end{equation}  
  where $\Omega_{d-2}$ is the volume of unit $(d-2)$-sphere. It is worth to note that the presence of power $\left[\alpha r^{-\left(d-1\right)\omega_{q}-d+3}\right]$ in Eq.~\eqref{exactsolution} modifies the asymptotic effect of  the quintessence term. Indeed, since  $-1<\omega_q<0$ \cite{Caldwell}, the quintessence term in Eq.~\eqref{exactsolution} has asymptotically flat spacetime behavior for $\omega_{q}>-\frac{d-3}{d-1}$,  while it becomes asymptotically dS-like for $\omega_{q}<-\frac{d-3}{d-1}$.  
  
  Thereafter, without loss of generality, we will consider only the  asymptotically dS behavior and set$\omega_{q}$ to the value $\omega_{q}=-\frac{d-2}{d-1}$ in numerical analysis.
  
  \section{Thermodynamics}\label{sec3}
  
  In this section, we will discuss the thermodynamics criticality for higher-dimensional charged AdS black hole surrounded by quintessence with cloud of strings present in extended phase space.  As usual, we treat the cosmological constant as a dynamical pressure of the black hole \cite{Kastor},
  \begin{equation}\label{pressurecosmolog}
  P=-\frac{\Lambda}{8\pi}.
  \end{equation}
  
  The Hawking temperature is related to the surface gravity via the formula $2\pi T=\kappa$, and is expressed as,
  
  \begin{equation}\label{temperature}
  T=\frac{1}{4\pi }\left[\frac{d \left(d-5\right)+6}{  \left(d-2\right) r_h} - \frac{\left(d-3\right) q^2}{r_h^{2 d-5}}-\frac{2 a}{ \left(d-2\right) r_h^{d-3}}+\frac{  \left(d-1\right)
  	\omega_q \alpha}{ r_h^{\omega_q \left(d-1 \right) +d-2}}+\frac{16 \pi   r_h }{ \left(d-2\right)}P\right],
  \end{equation}
  where the horizon $r_{h}$ is determined from the condition $f\left(r_{h}\right)=0$.  Using Bekenstein-Hawking formula \cite{Hawking}, the Entropy reads as,
  \begin{equation}\label{entrppy}
  S=\frac{A_{d-2}}{4}=\frac{\Omega_{d-2}}{4}r_{h}^{d-2}.
  \end{equation}
  From classical thermodynamics, the mass $M$ of the black hole is interpreted as the analogue of enthalpy rather than total energy of the spacetime  \cite{Kastor,Mann}. The generalized first law in the extended phase space  which account for the cosmological constant effect, cloud of strings and the quintessence contributions is then expressed as,
  \begin{equation}\label{firstlaw}
  dM=TdS+VdP+\Phi dQ+\mathcal{A} da+\mathcal{Q} d\alpha,
  \end{equation}
  where the conjugate quantities of the  parameters $P$, $Q$, $a$ and $\alpha$ read as:
  \begin{align}\label{conjugatequantities}
  V & =  \frac{\Omega _{d-2} r_h^{d-1}}{d-1},&  & \Phi  = \sqrt{\frac{2-d}{2 \left(3-d\right)}} \frac{q}{r_{h}^{d-3}},\\
  \mathcal{A} & = \frac{\Omega_{d-2}}{8 \pi } r_{h}, & & \mathcal{Q} =  \frac{\left(d-2\right) \Omega_{d-2}}{16\pi r_{h}^{\omega_q \left(d-1\right)}}.
  \end{align}
  It is worth to notice that the expressions of $V$ and $\Phi$ are similar to those of  $d$-dimensional RN-AdS black hole \cite{Mann}. All these quantities satisfy the following generalized Smarr formula:
  \begin{equation}\label{Smarrformula}
  M=\frac{d-2}{d-3} T S-\frac{2}{d-3}V P+\Phi Q+\left(\frac{d-1  }{d-3} \omega_q+1\right) \mathcal{Q} \alpha+ \frac{d-4}{d-3} \mathcal{A} a.
  \end{equation}
  
  Next, we focus on the study of criticality. To this end, we should first find out the stationary inflection points in the  $T(r_{h})-r_{h}$ diagram by solving the following equations:
  \begin{equation}\label{criticalpoint}
  \left.\frac{\partial T}{\partial r_{h}}\right|_{r_{c},P_{c}}=\left.\frac{\partial^2 T}{\partial r_{h}^2}\right|_{r_{c},P_{c}}=0.
  \end{equation}
  The critical point reads as,
  \begin{align}\label{criticalpointpc}
  \frac{16 \pi P_{c}}{\left(d-2\right) \left(d-3\right) }  =\frac{ 1}{r_{c}^2}-\frac{2 d-5 }{r_{c}^{2 d-4}}q^2 - \frac{2  a}{\left(d-2\right)r_{c}^{d-2}} 
  +\frac{ \left(d-1\right)^{2} \omega_q^{2}
  	+\left(d-1\right)  \left(d-2\right)\omega_q  }{\left(d-3\right) r_{c}^{-\left(d-1\right) (\omega_q +1)}}\alpha,
  \end{align}
   \begin{align}\label{criticalpointtc}
  T_{c}  =\frac{1}{4 \pi } \left[\frac{ 2d-6}{r_{c}}-\frac{\left(d-2\right) \left(2d-6\right)}{r_{c}^{2d-5}} q^2 -\frac{2  a}{r_{c}^{d-3}}+\frac{ \left(d-1\right)^2  (\omega_q +1)\omega_q }{r_{c}^{\left(d-1\right) \omega_q+d-2}}\alpha\right],
  \end{align}
  where the critical radius $r_{c}$ is the solution of equation:
   \begin{align}\label{criticalpointrc}
 \frac{a}{r_{c}^{d-4}} -
 \frac{\left(d-1\right)^2  (\omega_q +1) \left(\left(d-1\right) \omega_q +d-2\right) \omega_q}{2\left(d-3\right)r_{c}^{\left(d-1\right)\omega_q +d-3}}  \alpha
 +\frac{\left(d-2\right) \left(2 d-5\right)}{r_{c}^{2\left(d-3\right)}}  q^2 =1.
 \end{align}
  Looking at these equations, one can recover the result \footnote{We point out here a mistake in the calculation of critical point derived in \cite{Toledo2019}} for $d=4$  and $\omega_q=-2/3$:
  	 \begin{equation}\label{criticalpoint1}
  	r_c=\sqrt{\frac{6}{1-a}}Q\ ;\ 
  	P_c=\frac{\left(1-a\right)^{2}}{96\pi Q^{2}}\ ; \ T_{c}=\frac{\sqrt{6} (1-a)^{3/2}-9 \alpha  Q}{18 \pi  Q}.
  	\end{equation}
   
  Another key quantity characterizing the critical phenomena is the Gibbs free energy,
  \begin{equation}\label{Gibbsmts}
  G=M-TS.
  \end{equation}
  From Eq.~\eqref{ADMmass}, Eq.~\eqref{temperature} and Eq.~\eqref{entrppy}, one gets:
 \begin{align}\label{Gibbs}
 G=\frac{\Omega_{d-2}}{16 \pi} \left[r_h^{d-3}+\frac{2 d-5}{r_h^{d-3}} q^2-\frac{ 2d-6}{d-2}a r_h
 -\frac{\left(d-1\right) \omega_q +d-2}{r_h^{\left(d-1\right) \omega}}\alpha -\frac{16 \pi  P
 	r_h^{d-1}}{\left(d-1\right)\left(d-2\right)} 
 \right].
 \end{align}
  
\begin{figure}[h]
  	\begin{center}
  		\begin{tabbing}
  			\hspace{8.5cm}\=\kill
  			\includegraphics[scale=0.61]{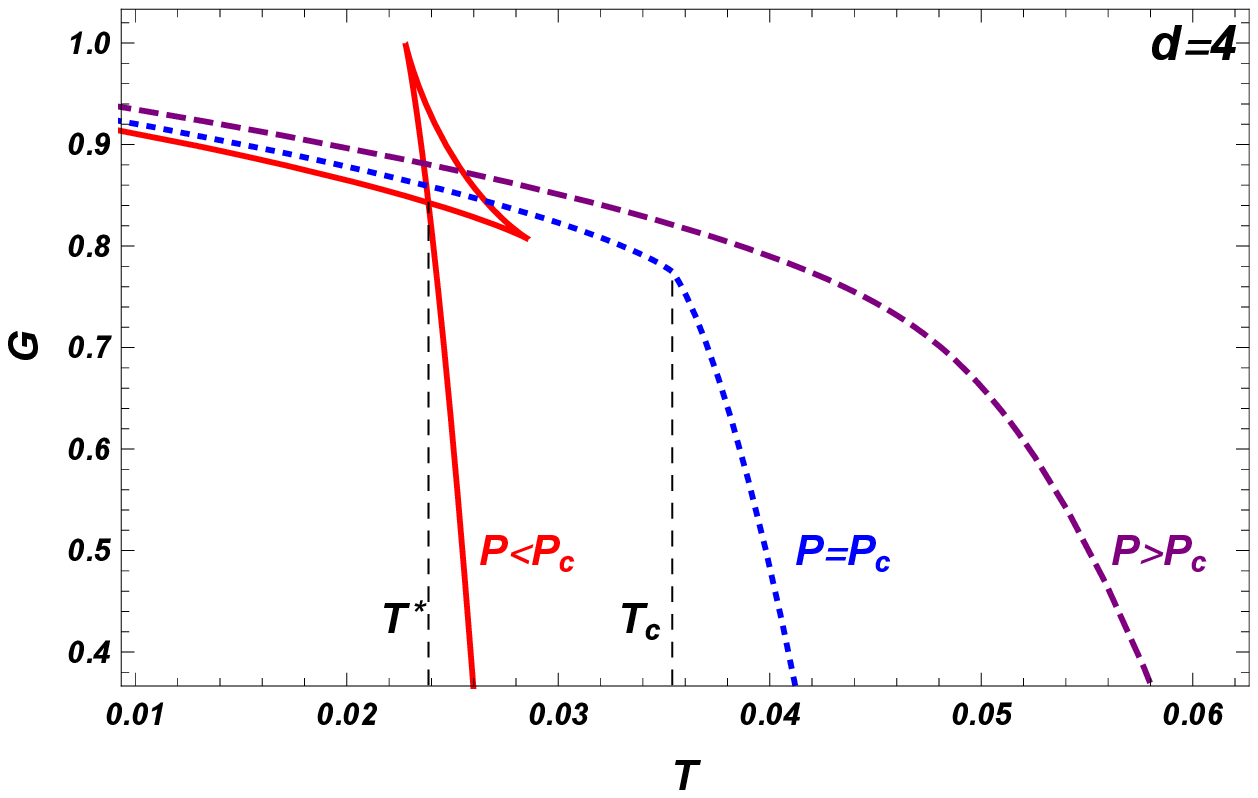}\>	\includegraphics[scale=0.61]{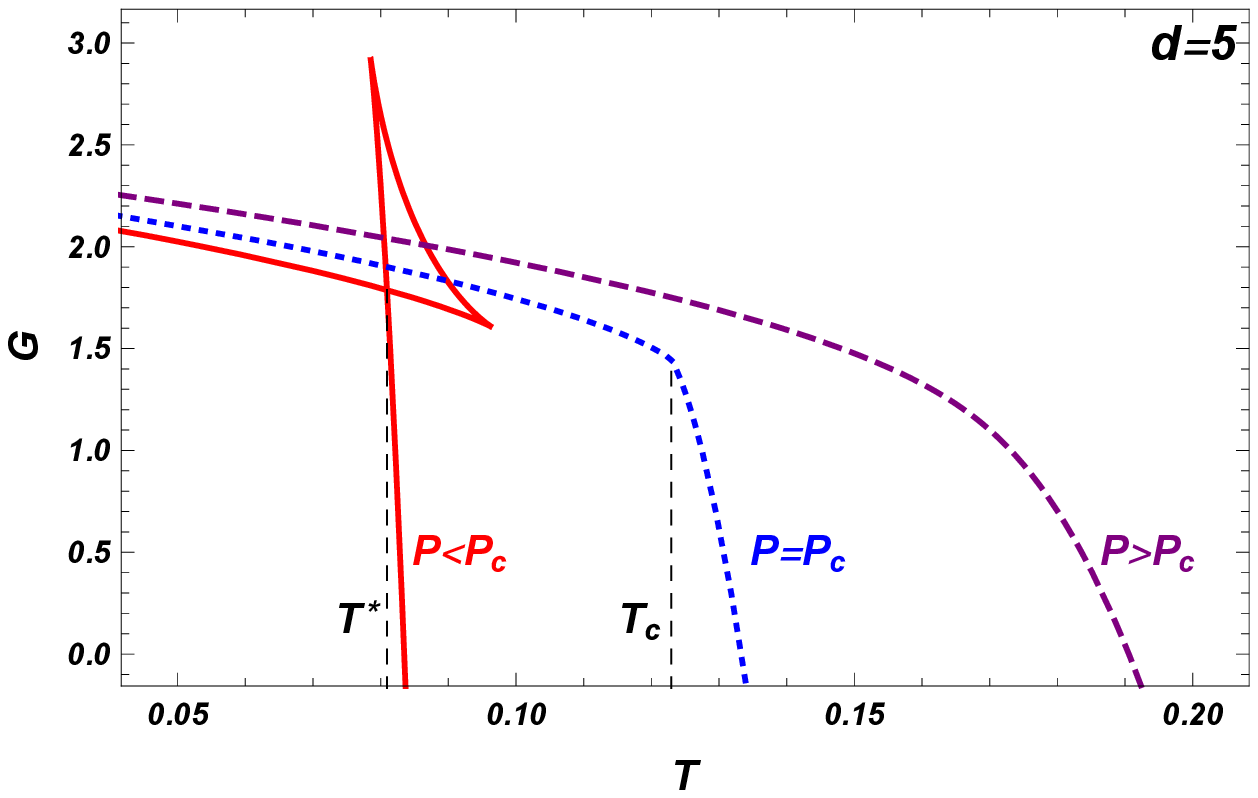}\\
  			\includegraphics[scale=0.61]{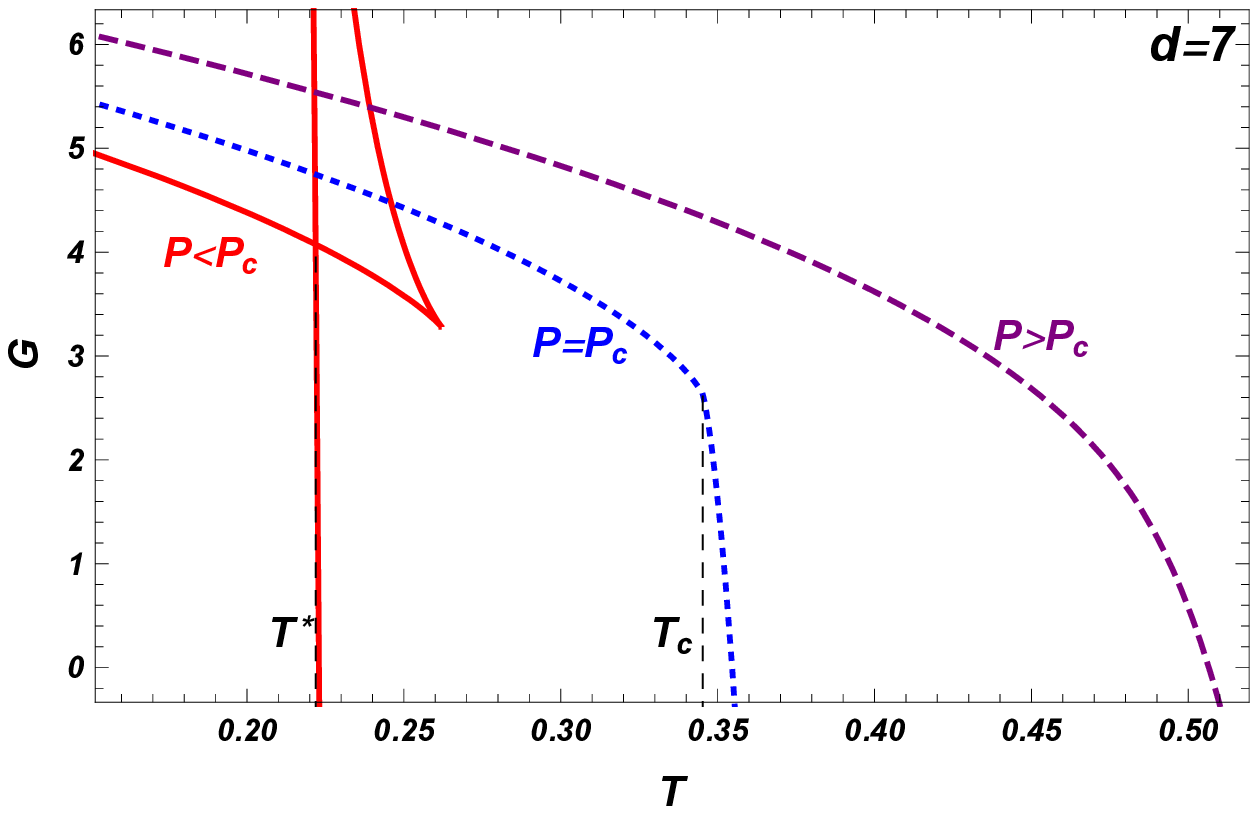}\>	\includegraphics[scale=0.61]{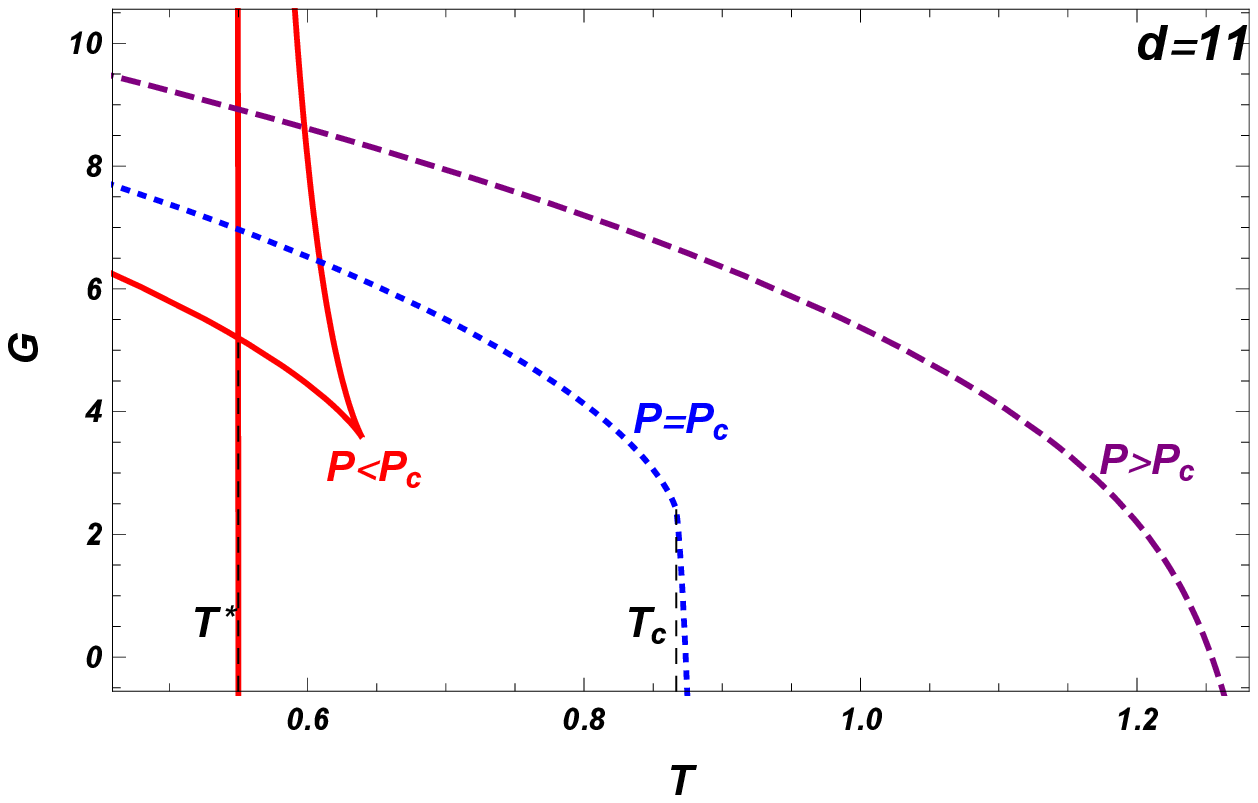}
  		\end{tabbing}
  	\end{center}
  	\caption{Gibbs free energy  as a function of temperature for fixed $q = 1$, $a=0.1$,  $\alpha=0.01$, $\omega_{q}=-\frac{d-2}{d-1}$, for various values of pressure and dimension.  } 
  	\label{figGT}
  \end{figure}

  Now, after calculating the main thermodynamics quantities, we turn our attention to the analysis of the corresponding phase transition. To this end, we plot in Fig.~\ref{figGT}, the variation of  Gibbs free energy as a function of the  temperature, for $d = 4$, $5$, $ 7$ and $11$. We see that the behavior for all dimensions is qualitatively similar, and  for $P<P_c$ the $G-T$ diagram develops a swallow tail behavior, revealing a first order phase transition as expected. This feature disappears above the critical pressure and thus no phase transition small/large black holes occurs. However, this phase transition transforms to a second order one at $P=P_{c}$.
  
\begin{figure}[h]
  	\begin{center}
  		\begin{tabbing}
  			\hspace{8.5cm}\=\kill
  			\includegraphics[scale=0.61]{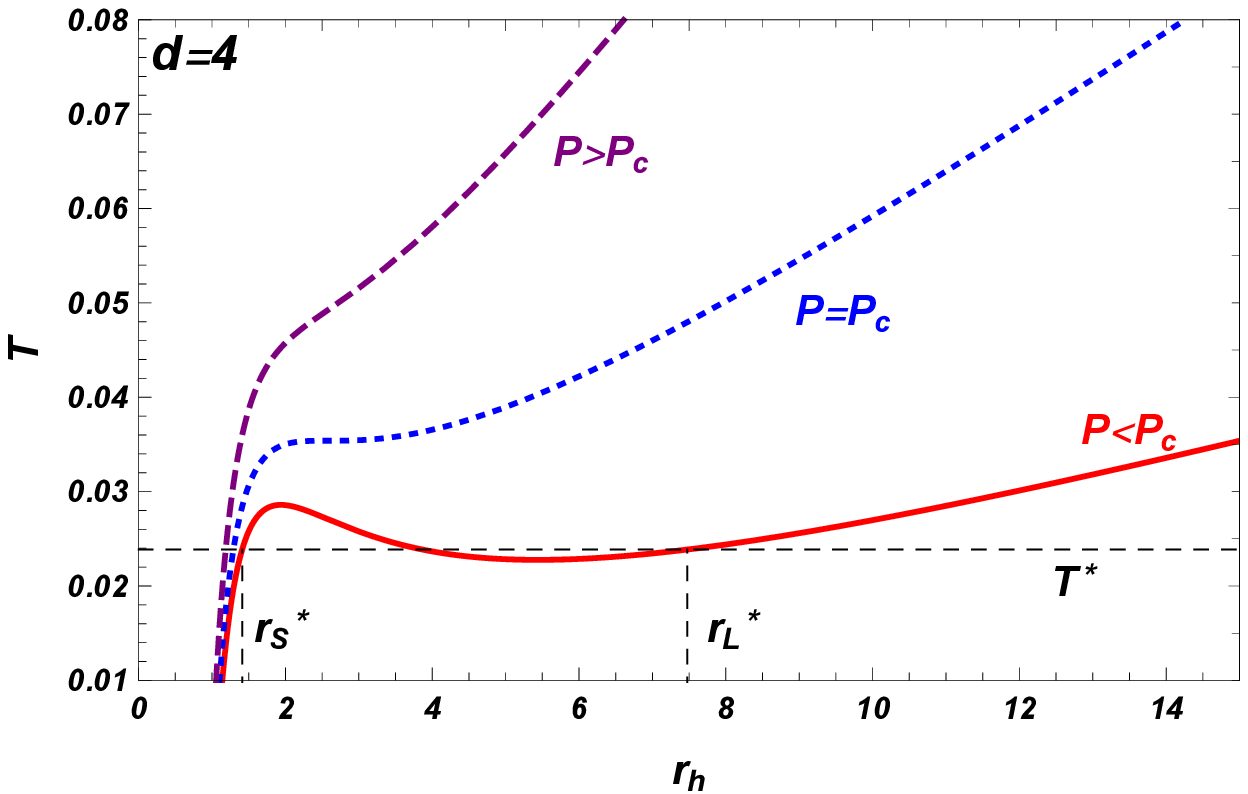}\>	\includegraphics[scale=0.61]{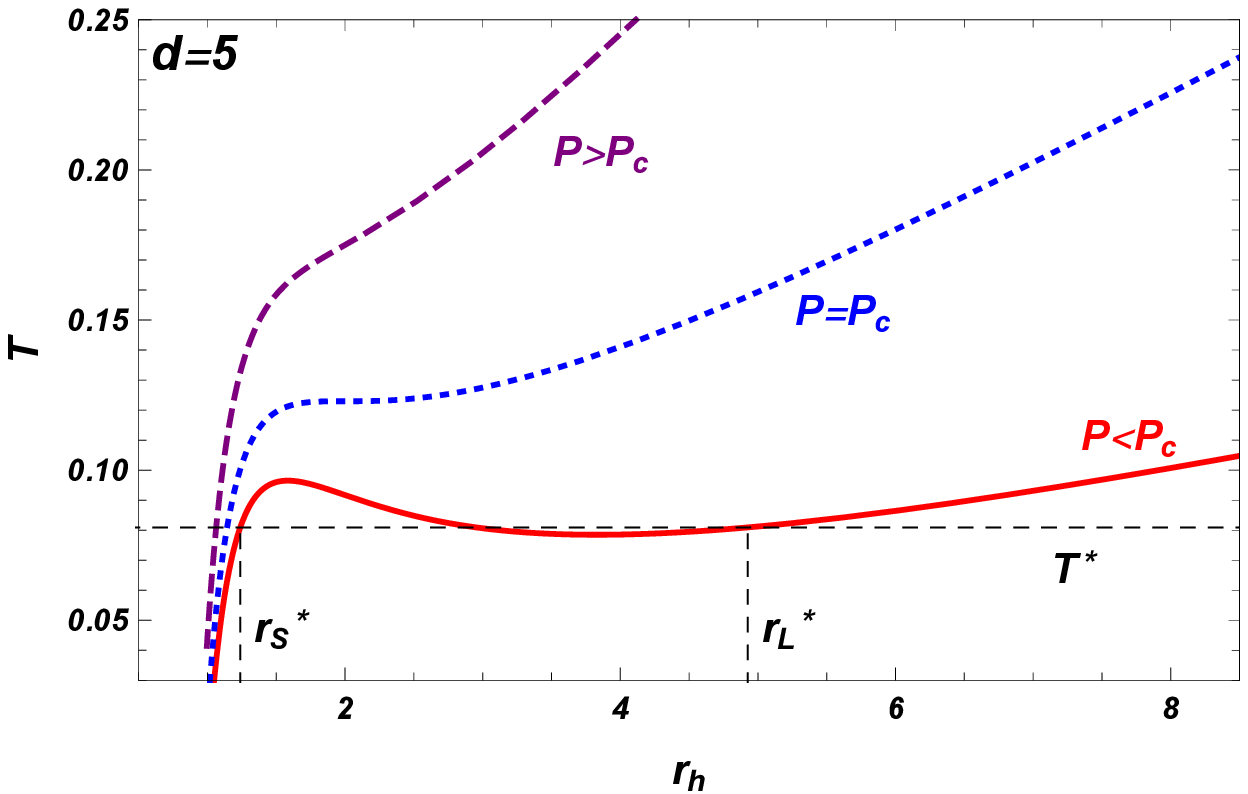}\\
  			\includegraphics[scale=0.61]{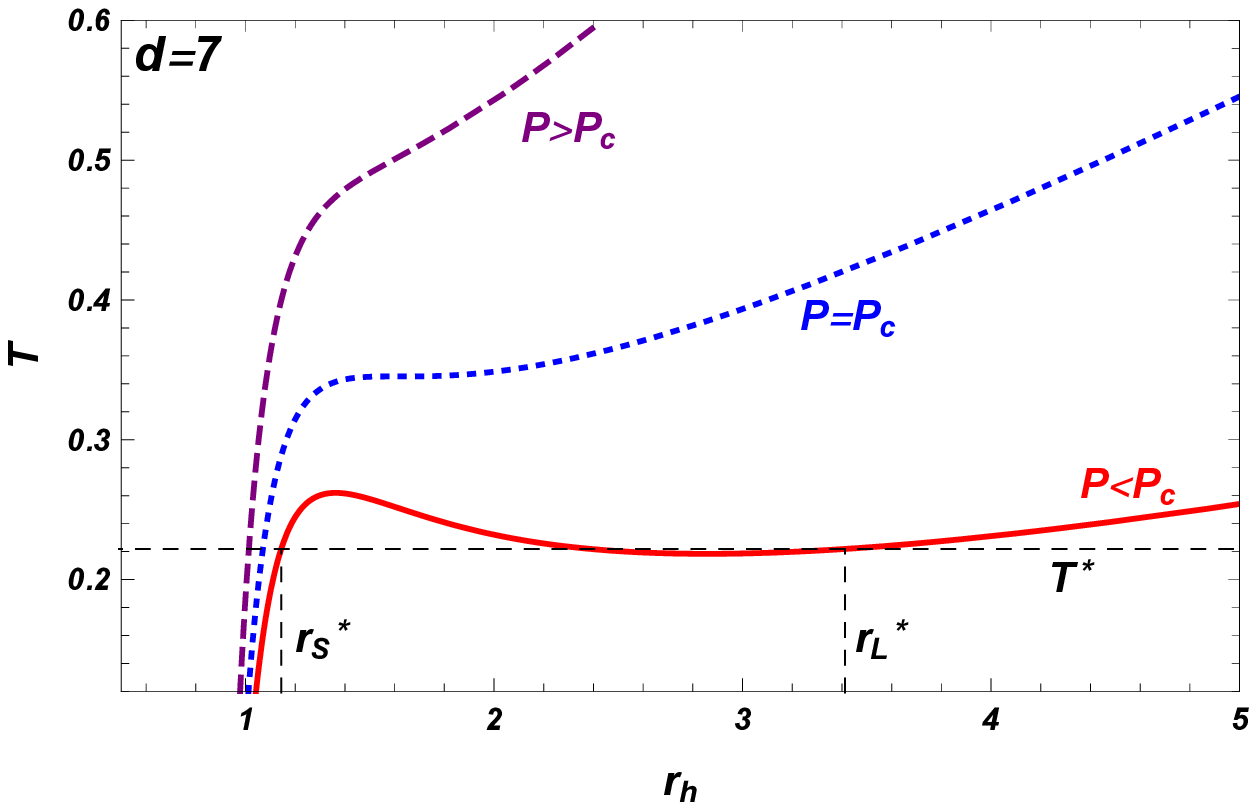}\>	\includegraphics[scale=0.61]{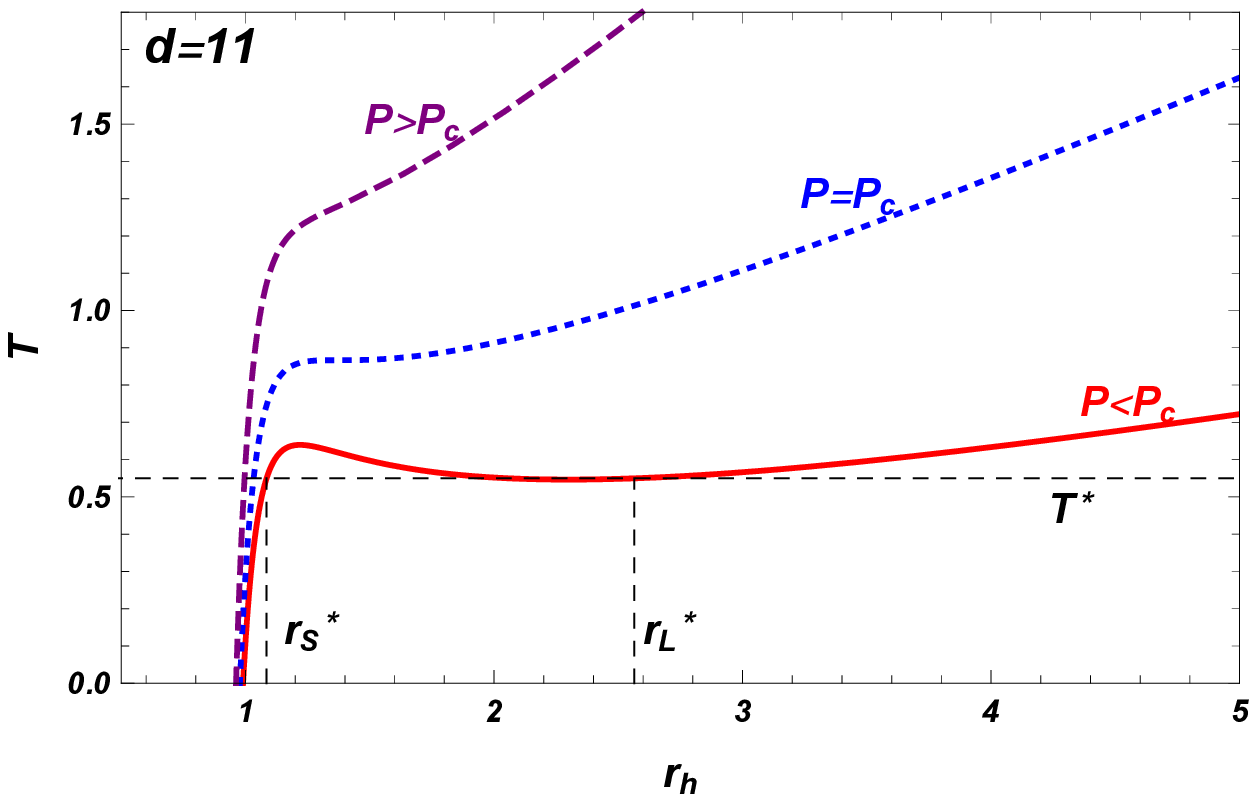}
  		\end{tabbing}
  	\end{center}
  	\caption{Hawking temperature  as a function of horizon  radius  for various pressures and dimensions.  We used: $q = 1$, $a=0.1$,  $\alpha=0.01$, $\omega_{q}=-\frac{d-2}{d-1}$.} 
  	\label{figTr}
  \end{figure}

  We also plot the temperature with respect to $r_h $ for $d = 4$, $5$, $7$ and $11$ in Fig.~\ref{figTr}, and find that, $T-r_h$ criticality exists for $P<P_c$. Therefore, the small/large black hole phase transition is disclosed at the point $\left(r^{*}_{S},T^{*}\right)$ and vanishes at  $\left(r^{*}_{L},T^{*}\right)$. Moreover, at the second order phase transition, we have $\left(r^{*}_{S},T^{*}\right)=\left(r^{*}_{L},T^{*}\right)=\left(r_{c},T_{c}\right)$. 
  
  According to Maxwell equal area law, to describe thermodynamics phase transition the oscillating part between $r^{*}_{S}(S^{*}_{S})$ and $r^{*}_{L}(S^{*}_{L})$ of the $T-r_h$ diagram should be replaced by an isotherm line, $T=T^{*}$, 
  \begin{equation}\label{areal law}
  \int_{S^{*}_{S}}^{S^{*}_{L}} TdS=T^{*}\left(S^{*}_{L}-S^{*}_{S}\right).
  \end{equation}
  This prescription reflects the fact that both phases have the same Gibbs free energy at the phase transition. To perform comparison, one can determine the critical point and the parameters describing  coexistence curve for different dimensions. The numerical results are summarized in Table \ref{numphtrans}.
  
 \begin{table}
 	\centering
 	\begin{tabular}{ccccccc}
 		\hline\hline 
 		$d$	& $T_{c}$ &$P_{c}$  & $r_{c}$  & $T^{*}$ & $r^{*}_{S}$ & $r^{*}_{L}$  \\ 
 		\hline 
 		4 & 0.03539 & 0.00268  & 2.58199 & 0.02385 & 1.41052  & 7.47307  \\ 
 		
 		5 &  0.12295 & 0.01952 & 1.99347 & 0.08094 & 1.23852 & 4.9265 \\ 
 		
 		6 &  0.22869 & 0.05768 & 1.75222 & 0.14824 & 1.17754 & 3.94063 \\ 
 		
 		7 & 0.34528 & 0.12143 & 1.6142 & 0.22188  & 1.14375 & 3.41249  \\ 
 		
 		11 & 0.86661 &  0.67705 & 1.37039 & 0.54974 & 1.08478 & 2.56388 \\ 
 		\hline
 		\hline 
 	\end{tabular} 
 	\caption{Critical point and parameters of two phases coexistence for fixed $q = 1$, $a=0.1$,  $\alpha=0.01$ and $\omega_{q}=-\frac{d-2}{d-1}$. $\left(T^{*}, r^{*}_{S},r^{*}_{L}\right)$ are obtained for $P/P_c = 0.4$} \label{numphtrans}
 \end{table}

  Now, we would like to describe the criticality in  extended phase space by exploring the behavior of heat capacity in the canonical ensemble  for fixed $q$ and $P$. Since, the local thermodynamics stability of the black holes is determined by the sign of heat capacity whose divergent point  may indicate  occurrence of phase transition between stable/unstable system. Thus, by using the  definition:
  \begin{equation}\label{Capacitydefinec}
  C=T\left. \dfrac{\partial S}{\partial T}\right| _{P,q}.
  \end{equation}
  After a straightforward calculation, using Eq.~\eqref{temperature}, \eqref{entrppy}, we obtain:
  \begin{equation}\label{Capacity}
  C_P=\frac{\left(d-2\right)\Omega_{d-2} r_h^{d -2}}{4 }\frac{A}{B},
  \end{equation}
  where $A$ and $B$ are given in the following simple forms,
  \begin{align}\label{numCapacity}
  A=  \frac{q^2}{r_h^{2d-6}} -\frac{2 a r_h^{4-d}}{\left(d-2\right)\left(d-3\right)} +\frac{16 \pi  P r_h^{2}}{\left(d-2\right)\left(d-3\right)}+\frac{ \left(d-1\right) \omega _q \alpha }{\left(d-2\right) r_h^{d+d\omega_{q}-\omega _q-3}}+1,
  \end{align}
  and
  \begin{align}\label{denomCapacity}
  B= \dfrac{ \left(2 d-5\right) q^2}{r_h^{2d-6}} + \frac{2 a r_h^{4-d}}{d-2}+\frac{16 \pi  P r_h^{2}}{\left(d-2\right)\left(d-3\right)}
  -\frac{
  	\left(	\left(d-1\right) \omega _q+ d-2\right)\left(d-1\right)\alpha \omega _q  }{\left(d-3\right) r_h^{d+d\omega_{q}-\omega _q-3}}-1.
  \end{align}
  
  We clearly see that, in addition to $P$ and $q$,  the heat capacity of the black hole is corrected by the parameters $\omega_{q}$, $\alpha$, $a$ and the  spacetime dimension. Their corresponding terms conspire,  in the denominator B, so that $C_P$ diverges. 
  
\begin{figure}[h]
  	\begin{center}
  		\begin{tabbing}
  			\hspace{8.5cm}\=\kill
  			\includegraphics[scale=0.61]{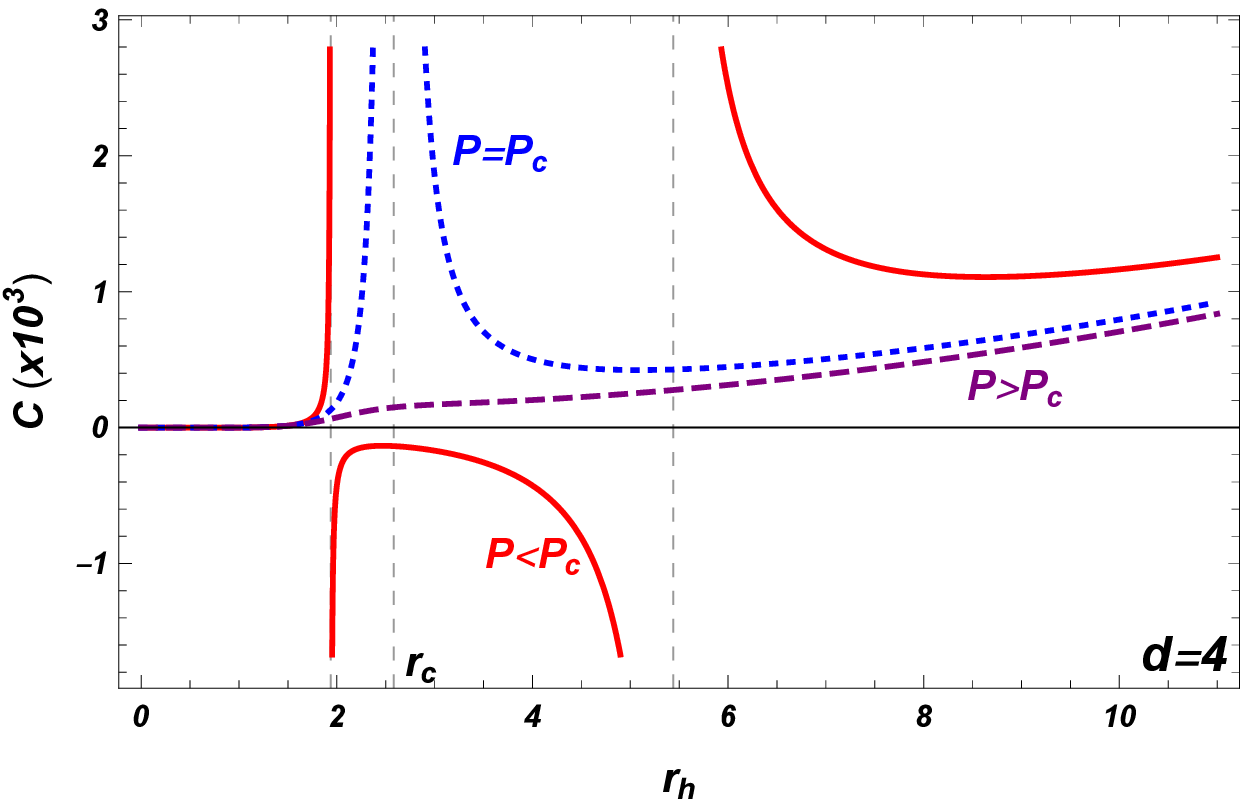}\>	\includegraphics[scale=0.61]{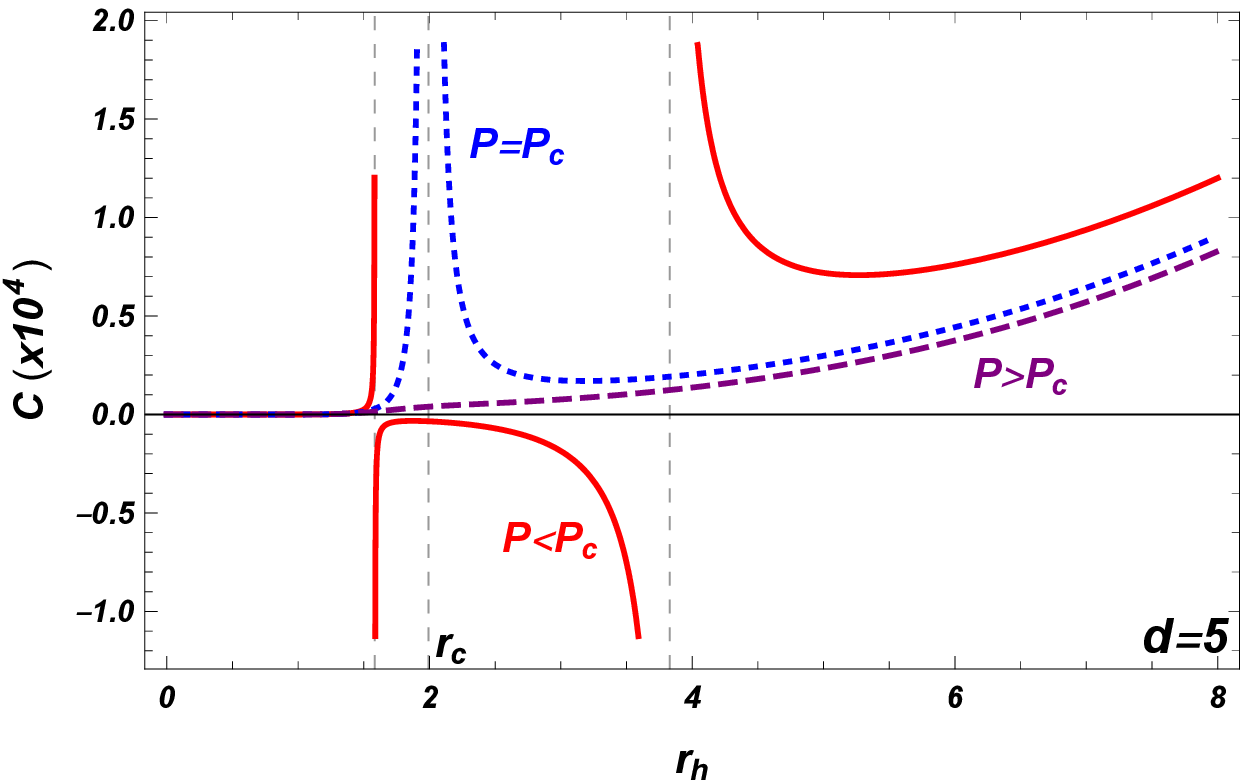}\\
  			\includegraphics[scale=0.61]{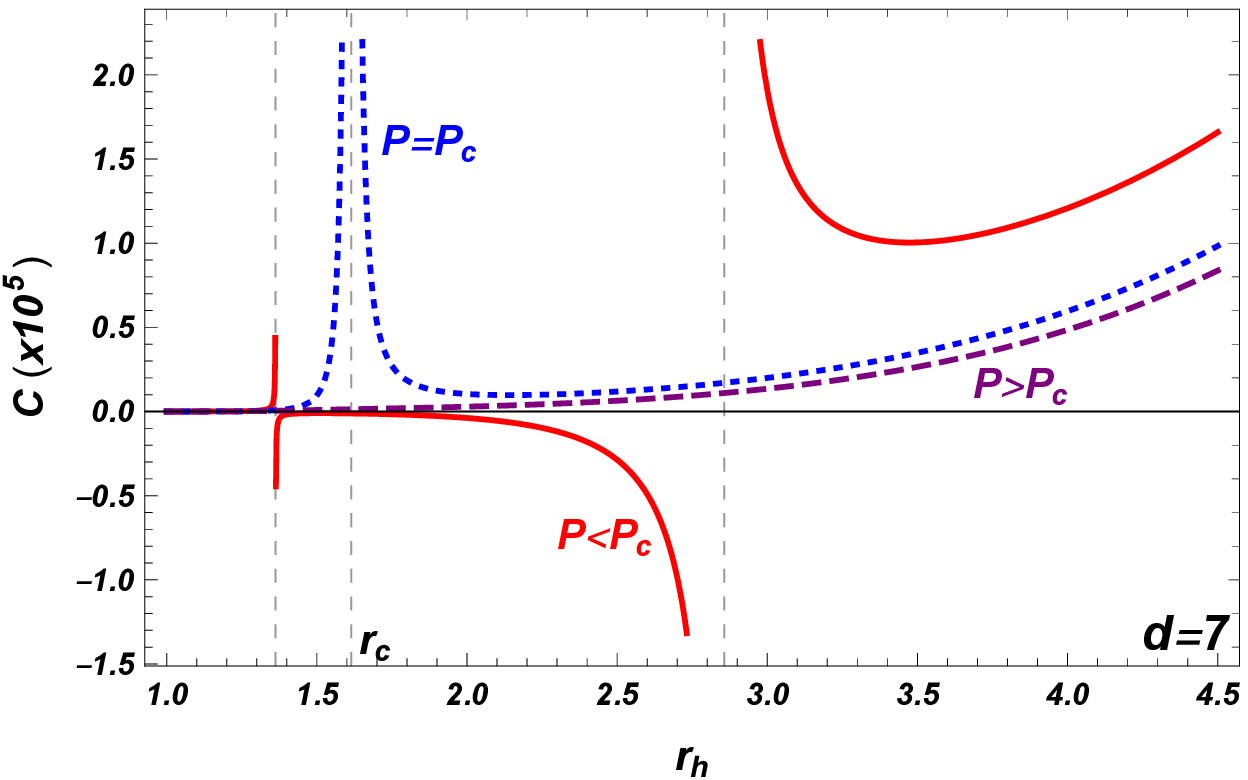}\>	\includegraphics[scale=0.61]{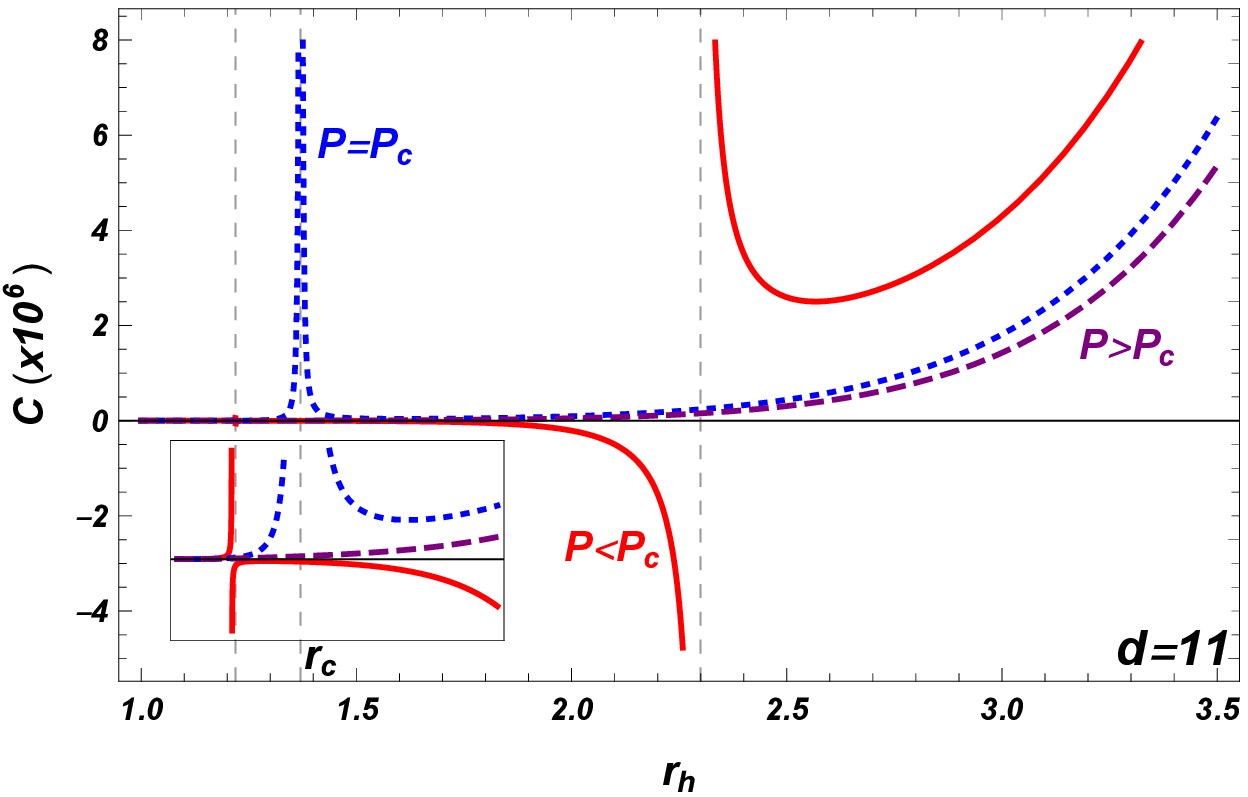}
  		\end{tabbing}
  	\end{center}
  	\caption{The behavior of heat capacity versus $r_h$  for different dimensions. Here, we use $q = 1$, $a=0.1$,  $\alpha=0.01$, $\omega_{q}=-\frac{d-2}{d-1}$.} 
  	\label{figCpr}
  \end{figure}

  Since the analytic investigation of the heat capacity, Eq.~\eqref{Capacity} is not straightforward, we resort to numerical analysis. Indeed Fig.~\ref{figCpr} illustrates the behavior of $C_P$ for different dimensions. 
  It is shown that the heat capacity is a negative discontinuous function at two separated points when $P<P_c$ (red lines), hence the system is thermodynamically unstable in the region delimited by the two divergent points. However stability is restored outside this region. Once, the second order phase transition takes place (blue dashed lines),  the non-stability disappears and all phases recover stability, except at the critical point $r_c$ where the heat capacity tends again to infinity.  Above the critical pressure (dashed purple lines), no divergent points for $C_P$ survive and the phase transition disappears. Note here that this feature is verifiable for all dimensions, and that the unstable region reduces as we gradually increase $d$.
  
  \section{Thermodynamics geometry}\label{sec4}
  Study of black hole phase transitions from the point of view of thermodynamic geometry has been quite active recently \cite{Chabab:2019mlu}. Hence,  in this section, we will apply geothermodynamics concept to describe the thermodynamics phase transition of higher dimensional charged AdS black hole surrounded by quintessence with strings cloud  background. The Legendre transformation used to build our metric is that given by Quevedo \cite{Quevedo} as,
  \begin{equation}\label{Quevedo}
  ds_{Quv}^{2}= \left(S\frac{\partial M}{\partial S}+Q\frac{\partial M}{\partial Q}\right) \left(-\frac{\partial^{2} M}{\partial S^{2}}dS^{2}+\frac{\partial^{2} M}{\partial Q^{2}}dQ^{2}\right).
  \end{equation}
  In order to obtain the curvature singularity of the Quevedo metric, we calculate the Ricci scalar.  Analytically, the lengthy  obtained results can be reduced to the simple form:
  \begin{equation}\label{scalarQuevedo}
  R^{Q}= \dfrac{\frac{\partial^{2} M}{\partial S^{2}} \left(\frac{\partial M}{\partial Q}\right)^{2}-\frac{\partial^{2} M}{\partial Q^{2}} \left(\frac{\partial M}{\partial S}\right)^{2}}{\left(Q \frac{\partial M}{\partial Q}+S \frac{\partial M}{\partial S} \right)^{3} \left(\frac{\partial^{2} M}{\partial S^{2}} \frac{\partial^{2} M}{\partial Q^{2}}\right)}.
  \end{equation}
  
\begin{figure}[h]
 	\begin{center}
 		\begin{tabbing}
 			\hspace{8.5cm}\=\kill
 			\includegraphics[scale=0.61]{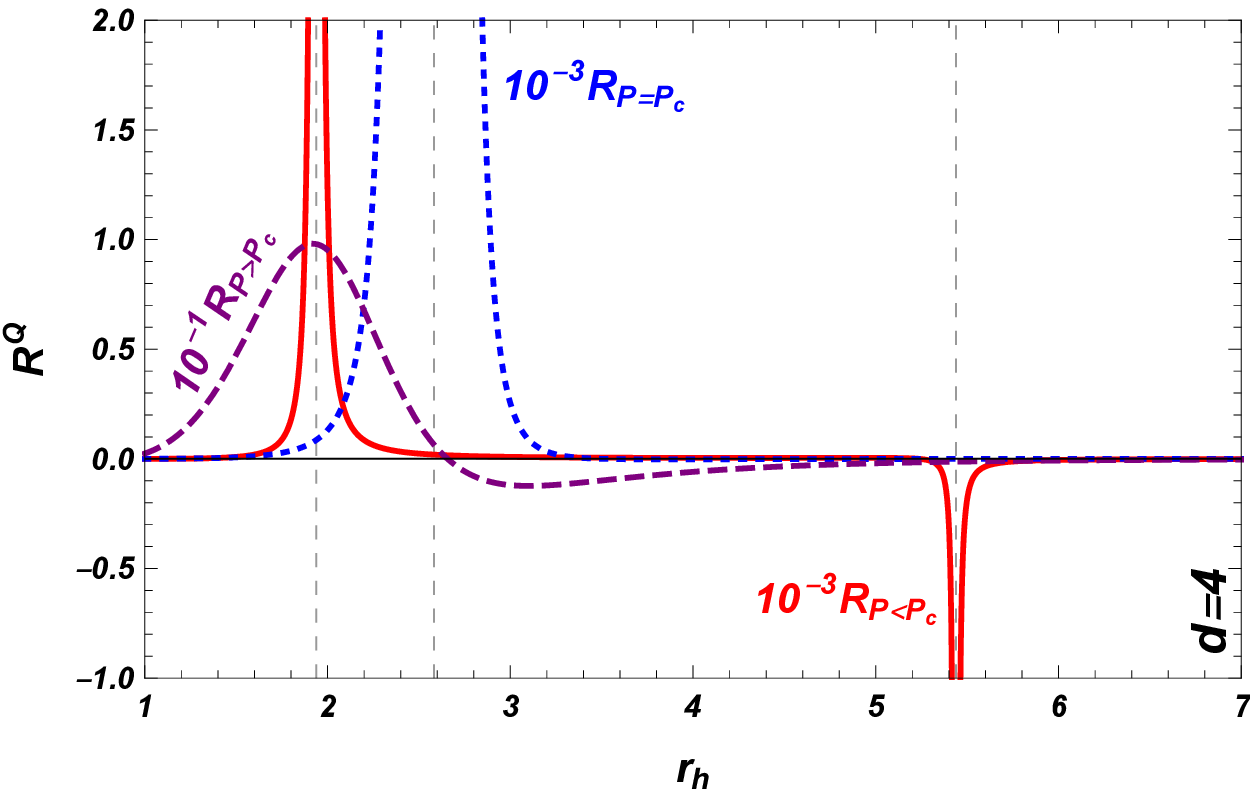}\> \includegraphics[scale=0.61]{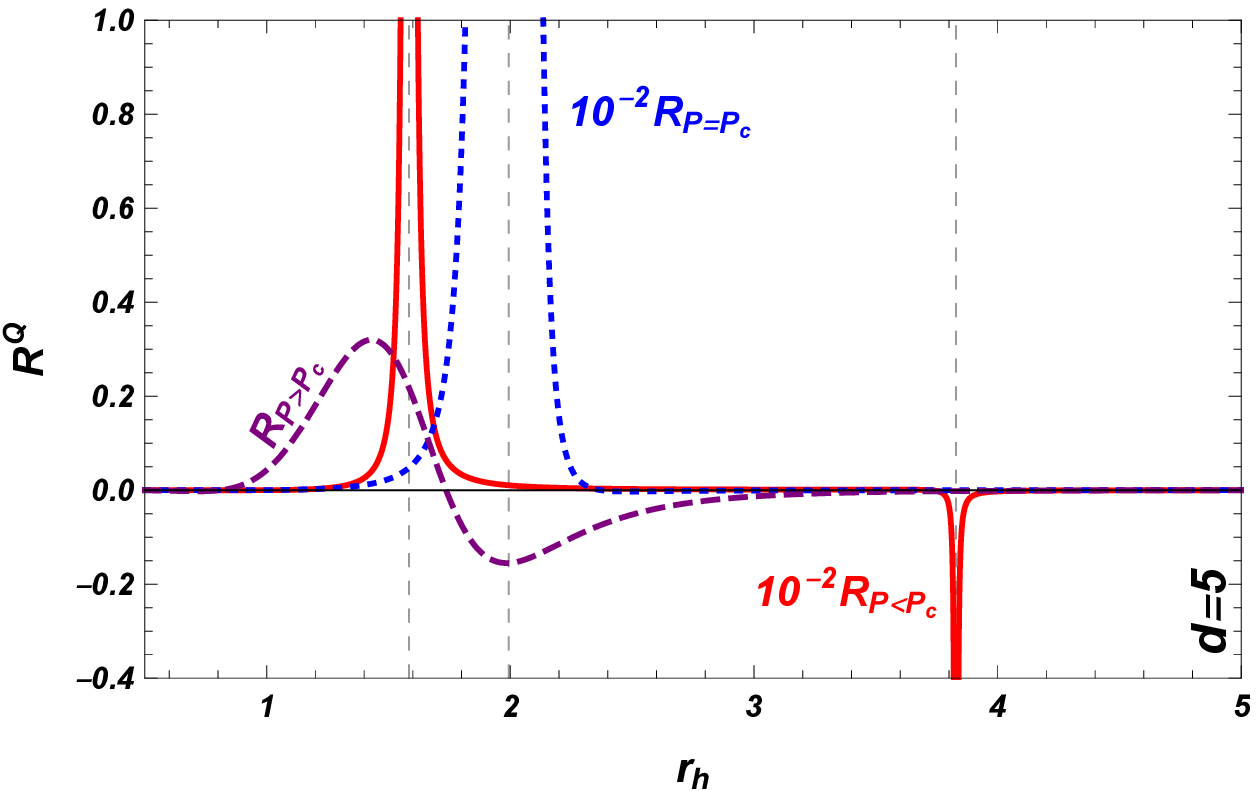}\\
 			\includegraphics[scale=0.61]{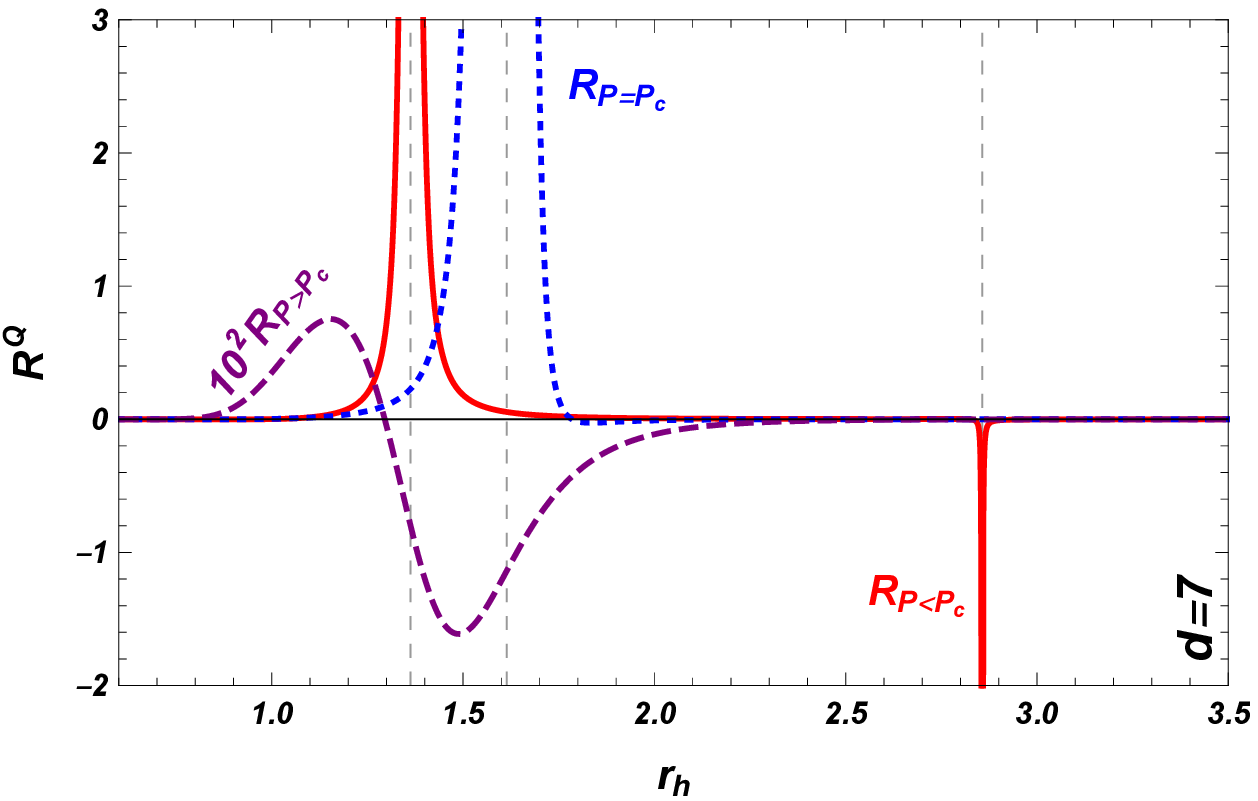}\>	\includegraphics[scale=0.61]{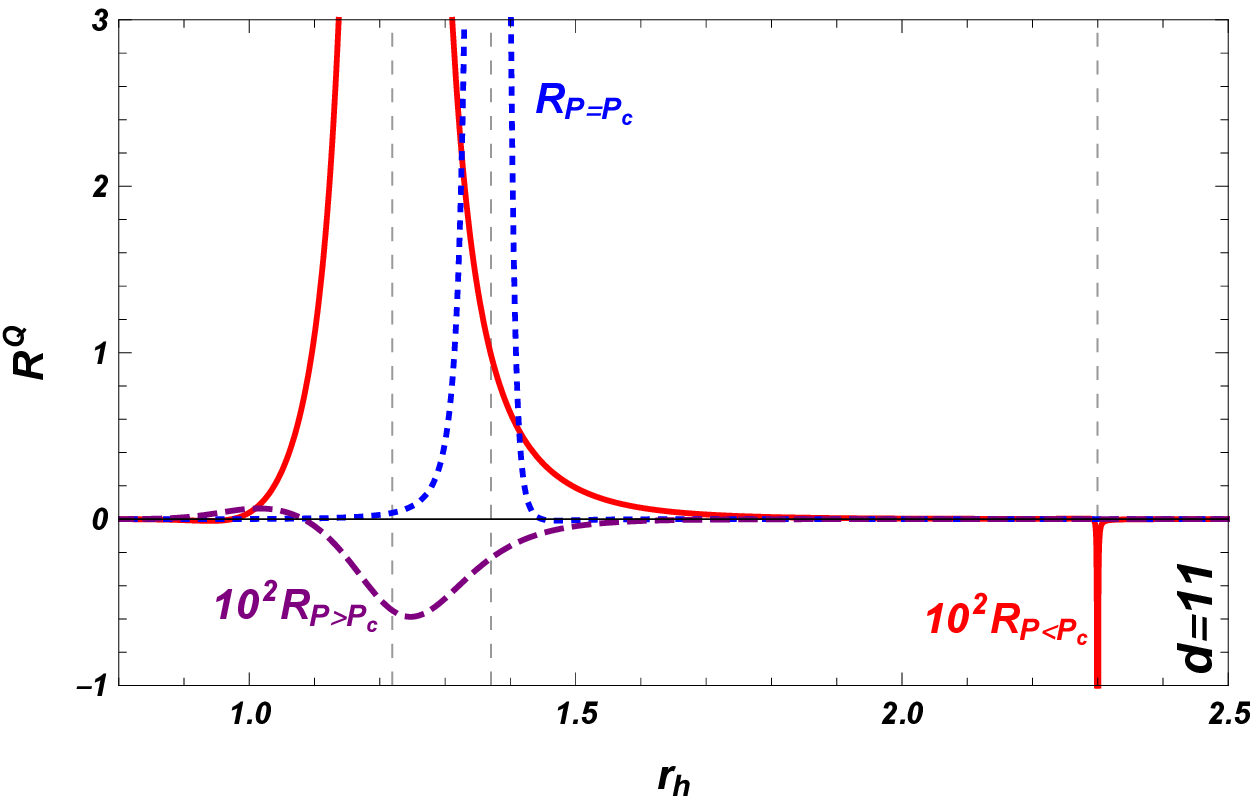}
 		\end{tabbing}
 	\end{center}
 	\caption{The Quevedo curvature $R^{Q}$ versus $r_h$ for different dimensions, with $q = 1$, $a=0.1$,  $\alpha=0.01$, $\omega_{q}=-\frac{d-2}{d-1}$, for various pressures and dimensions.} 
 	\label{RQuv}
 \end{figure}

  Since we are looking for the divergent points of Ricci scalar $R^{Q}$, we  focus on the denominator of $R^{Q}$ given by the last two terms in  Eq.~\eqref{scalarQuevedo}. Thus the divergent points occur when,  
  \begin{equation}\label{divscalarQuevedo1}
  Q \frac{\partial M}{\partial Q}+S \frac{\partial M}{\partial S} =0,
  \end{equation}
  or
  \begin{equation}\label{divscalarQuevedo2}
  \frac{\partial^{2} M}{\partial S^{2}} \frac{\partial^{2} M}{\partial Q^{2}}=0.
  \end{equation}
  The  scalar curvature of Quevedo is computed and plotted in Fig.~\ref{RQuv}. From Figs.~\ref{figCpr}, \ref{RQuv} we clearly see that the divergent points of Quevedo metric are fully consistent with the result obtained from the heat capacity. Indeed, there are two singular points above the critical pressure that coincide with those of the heat capacity, while for $P=P_c$  we get only one singular point at  $r_c$.  The singularities gradually disappear when the phase transition fades out away. Hence, the Legendre invariant of Quevedo is appropriate to predict  the thermodynamics criticality of  higher dimensional RN AdS black holes surrounded with quintessence and cloud of string backgrounds. 

  \FloatBarrier
  \section{Behavior near critical point}\label{sec5}
  Here, we follow a similar treatment as \cite{Li} for $a=0$ and $d=4$, and  \cite{Kubiznak}  for RN-AdS black holes. Our objective is to calculate the critical exponents $\xi,\ \beta, \ \gamma, \ \delta$ describing the behavior of relevant thermodynamics quantities near the critical point found in section \ref{sec3}. To this end, it is appropriate to introduce first the following notations,
  \begin{equation}\label{reduit}
  t=\frac{T-T_{c}}{T_{c}}, \ \ \epsilon=\frac{r_{h}-r_{c}}{r_{c}}, \ \  p=\frac{P}{P_{c}}.
  \end{equation}
  The critical exponents can be expressed in terms of a power law around the critical temperature as:
  \begin{equation}\label{exponet}
  C_{V} \propto  \left| t \right|^{-\xi}, \ \ \ \eta  \propto  \left| t \right|^{\beta},\ \ \ \kappa _{T} \propto  \left| t \right|^{-\gamma}, \ \ \  \left| P-P_{c} \right| \propto  \left| \epsilon \right|^{\delta}.
  \end{equation} 
  Since the entropy $S$ in Eq.~\eqref{entrppy} do not dependent on the temperature $T$, then 
  \begin{equation}\label{specificheatv}
  C_{V} = T\left|\frac{\partial S}{\partial T}\right|_{V}=0.
  \end{equation} 
  Therefore,  the critical exponent $\xi$ characterizing the behavior of the specific heat with fixed volume is  simply $\xi=0$.  Using Eq.~\eqref{temperature},  \eqref{criticalpointpc}, \eqref{criticalpointtc}, \eqref{criticalpointrc} and \eqref{reduit}, we can expand the equation of state near the critical point ($\left| t \right|\ll1$,  $\left| \epsilon \right|\ll1$),
  \begin{equation}\label{serie}
  p = 1+b_{10}t+b_{01}\epsilon+b_{11}t \epsilon +b_{02} \epsilon^{2} + b_{03} \epsilon^{3}+\mathcal{O} \left(t \epsilon^{2},\epsilon^{4}\right),
  \end{equation} 
  where
  \begin{equation}\label{b00b01}
  b_{01}=b_{02}=0,
  \end{equation} 
  \begin{equation}\label{b10b11}
  b_{10}=-b_{11}=\frac{d-2}{4} \frac{T_{c}}{P_{c} r_{c}},
  \end{equation} 
  \begin{multline}\label{coeifserie}
  b_{03}=\frac{d-2}{96 \pi P_{c} r_{c}}\left[\frac{12 \left(d-3\right)}{r_c}- \frac{2 \left(d-3\right) \left(d+2\right)}{r_c^{d-3}} a -\frac{4  d \left(d-3\right) \left(d-2\right) \left(2 d-5\right) }{r_c^{2 d-5}} q^2 
  \right.\\
  \left.
  +\frac{  \omega _q   \left(d-1\right)^2 \left(\omega _q+1\right) \left[\left(d-1\right) \omega _q+d-2\right]
  	\left[\left(d-1\right) \omega _q+d+3\right]}{r_c^{d \left(\omega _q+1\right)-\omega _q-2}}\alpha
  \right].
  \end{multline} 
  Clearly, Eq.~\eqref{serie} has a  similar shape to that of the RN-AdS black holes derived in \cite{Kubiznak}, then one would  expect to get the same critical exponents $\beta$, $\gamma$ and $\delta$. Using Maxwell's equal area law, we obtain the following two equations,
  \begin{equation}\label{maxwelarea}
  b_{11}t \left(\epsilon_{s}-\epsilon_{l}\right) + b_{03} \left(\epsilon^{3}_{s}-\epsilon^{3}_{l}\right)=0,
  \end{equation} 
  \begin{equation}\label{intmaxwelarea}
  0=\int_{\epsilon_{s}}^{\epsilon_{l}} \epsilon \frac{d p}{d \epsilon} d \epsilon ,
  \end{equation} 
  where $\epsilon_{s}$ corresponds to the small black hole and $\epsilon_{l}$ to the large one. Solving these equations, we get:
  \begin{equation}\label{epsilon}
  \epsilon_{l}=-\epsilon_{s}=\sqrt{-\frac{b_{11} t}{b_{03}}},
  \end{equation} 
  \begin{equation}\label{etaexponent}
  \eta\propto r_{hl}-r_{hs}=2 r_{c} \sqrt{-\frac{b_{11} t}{b_{03}}}.
  \end{equation} 
  Hence, $\beta$ exponent describing the behavior of $\eta$ is equal to $\beta=1/2$. To derive the critical exponent $\gamma$  which characterize the  isothermal compressibility $\kappa_T$ comports, we use Eq.~\eqref{serie} and the volume $V$ defined in Eq.~\eqref{conjugatequantities},  
  \begin{equation}\label{kappaexponent}
  \kappa_T=-\frac{1}{V}\left.\frac{\partial V}{\partial P}\right|_{T}\propto \frac{\partial \epsilon}{\partial p}=\left(1/b_{11}\right) t^{-1},
  \end{equation} 
  and get $\gamma=1$. Finally, for the exponent $\delta$,  it is easily obtained by taking $t=0$ in Eq.~\eqref{serie},
  \begin{equation}\label{delata}
  p-1=b_{03} \epsilon^{3},
  \end{equation} 
  which yields $\delta=3$.
  
  To summarize the critical exponents derived in this work are exactly the same as those obtained in many papers, such as: Mann et al. paper dealing with charged AdS black holes in 4d and $d$ spacetime dimensions \cite{Kubiznak, Mann}; non-linear charged black hole considered in \cite{Nam};  RN-AdS BH surrounded by quintessence \cite{Li},  and  also \cite{Toledo2019} with cloud of strings.


  \section{Effects of the cloud of strings  and quintessence on the  criticality of uncharged  black hole} \label{sec6}
  Kubiznak and Mann \cite {Kubiznak} have shown that the electrical charge $q$ of black holes plays an important role in the Van der Waals analogy. However,  it has been shown that small/large black hole phase transition occurs even for uncharged  black hole class, if some conditions are satisfied. For instance, Gauss-Bonnet black holes in AdS space~\cite{Cai2013}, Einstein-Gauss-Bonnet (EGB) black holes surrounded by a cloud of strings~\cite{Ghaffarnejad}.  Our objective here is to check if the absence of electrical charge modifies our findings derived in previous sections .
  
  \subsection{Four-dimensional AdS-Schwarzschild  black hole}
\begin{figure}[h]
  	\begin{center}
  		\begin{tabbing}
  			\hspace{8.5cm}\=\kill
  			\includegraphics[scale=0.61]{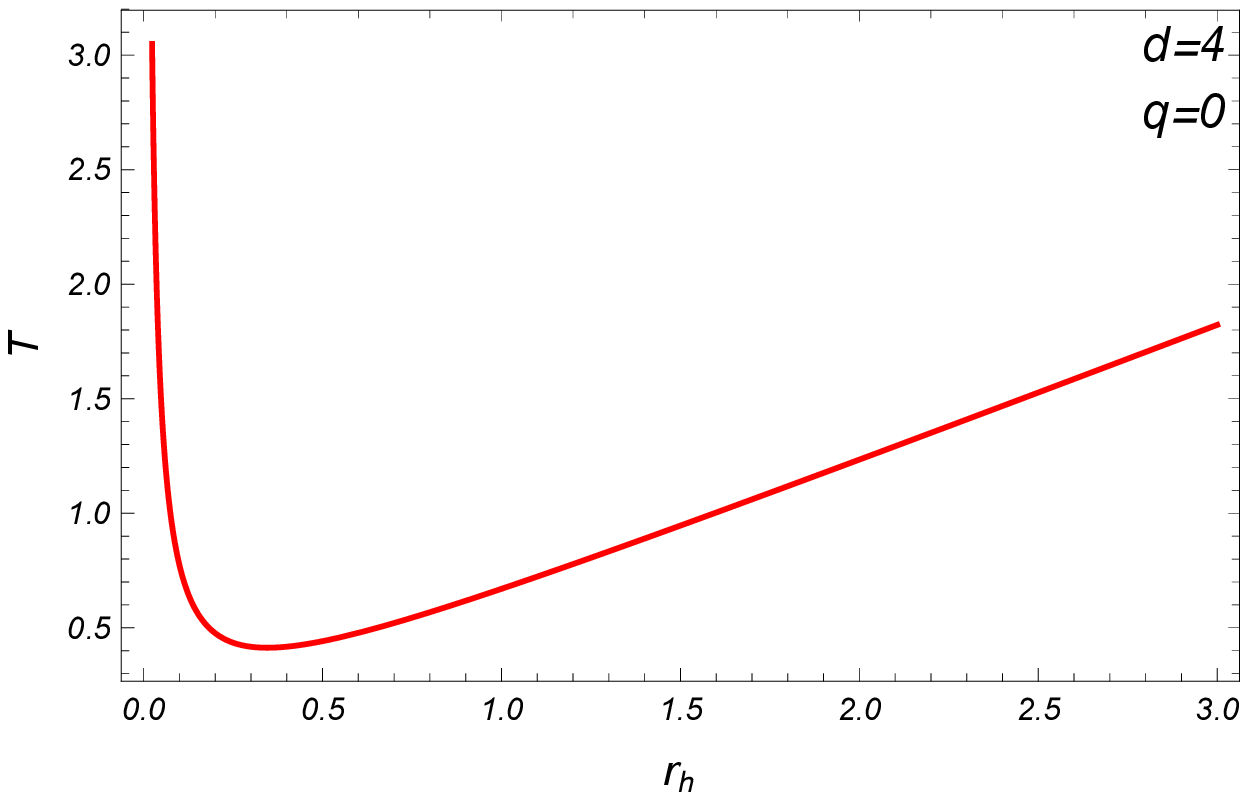}\> \includegraphics[scale=0.63]{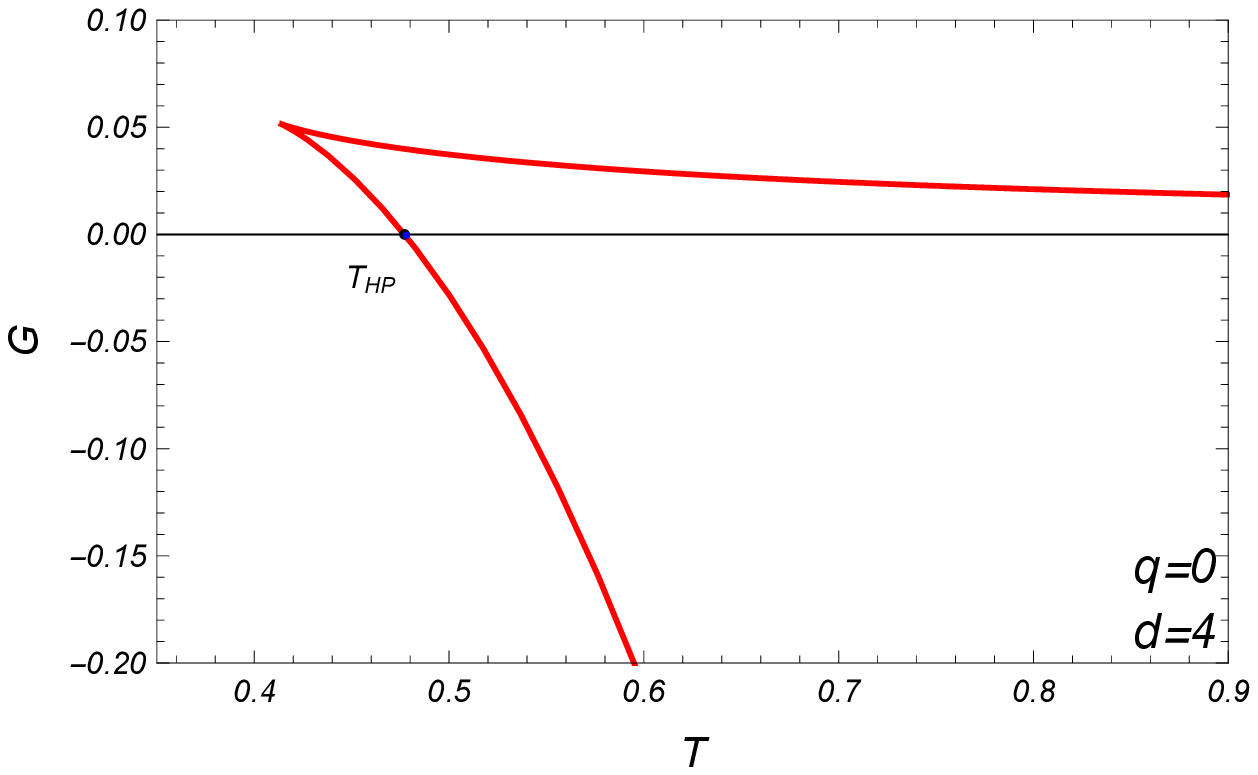}
  		\end{tabbing}
  	\end{center}
  	\caption{Right: The Gibbs free energy as a  function of the temperature. Left: Hawking temperature as a function of horizon  radius  for AdS-Schwarzschild  black hole in $d=4$, with $a=0.1,\ \alpha=0.01,\ \omega_q=-2/3$. } 
  	\label{q0d4}
  \end{figure}

  By fixing $d=4$  the equation~\eqref{criticalpoint} for the critical point reduces to:
   \begin{align}\label{criticalpointuncharged}
   &	r_c  = \left[\frac{9\, \alpha\,  \omega_q\, \left(3\, \omega _q^2+5\, \omega _q+2\,\right)}{2\,\left(a-1\right)}\right]^{\frac{1}{3\,\omega _q+1}}; \\
   &	 P_c =
  	\frac{\left(3\, \omega _q+1\right)\, \left(1-a\right)}{24\, \pi\, \left(1+\omega_q\right)\, r_{c}^{2}}; \\
  	&	T_c=
  	 \frac{\left(3\, \omega _q+1\right) \left(1-a\right)}{2\, \pi\, \left(2+3\omega_q\right)\, r_{c}}.
  \end{align}
  To ensure positivity of the  critical pressure and radius, the following conditions should be fulfilled, 
  \begin{align}\label{condition1}
  	a>1, \qquad -1<\omega_q<-2/3. 
  \end{align}
   These constraints  lead instead to a negative temperature,  therefore  the black hole cannot show a thermodynamic behavior similar to that of Van der Waals fluid or RN-AdS BH.  However, if $a<1$, it behaves thermally as AdS Schwarzschild black hole as shown in Fig.~\ref {q0d4}, thus only the Hawking-Page phase transition subsists.

\FloatBarrier
  \subsection{AdS-Schwarzschild   black hole with $a\ne 0$, $\alpha=0$  and $d \ge 5$}
\begin{figure}[h]
 	\begin{center}
 		\begin{tabbing}
 			\hspace{8.5cm}\=\kill
 			\includegraphics[scale=0.61]{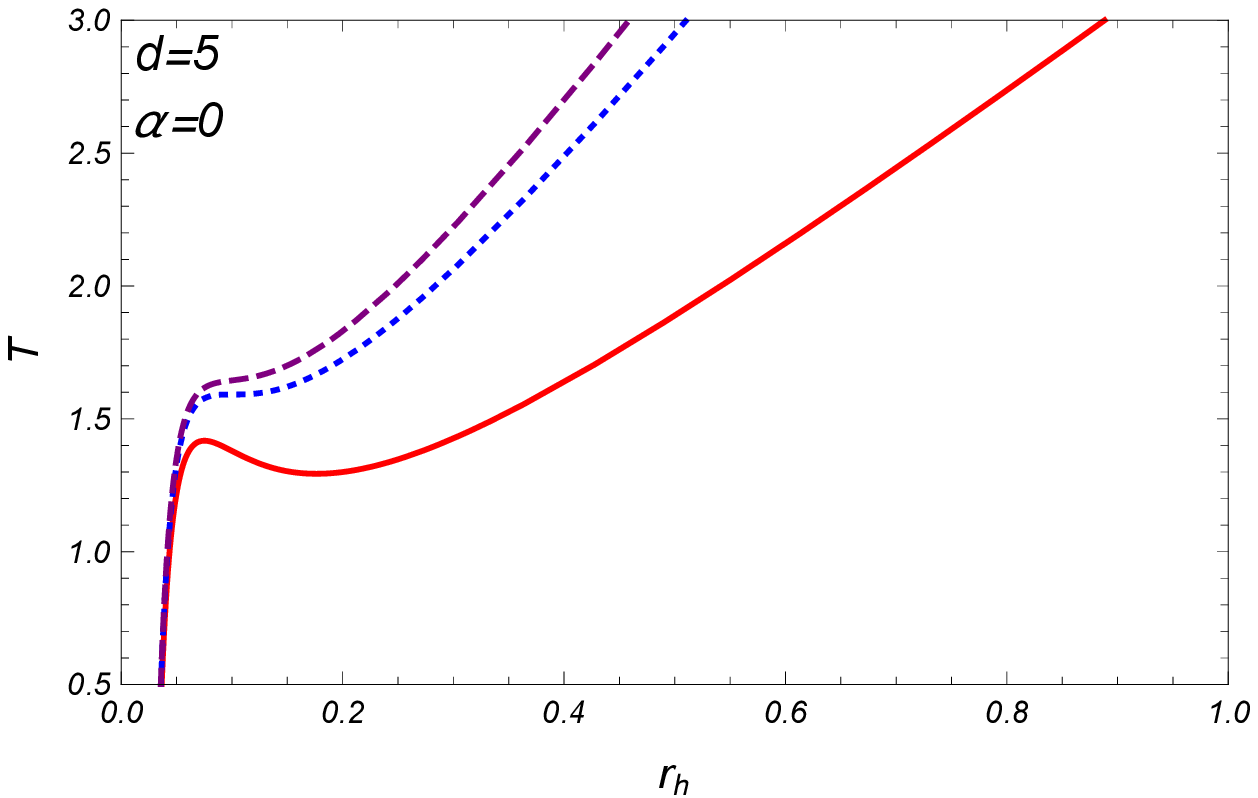}\> \includegraphics[scale=0.625]{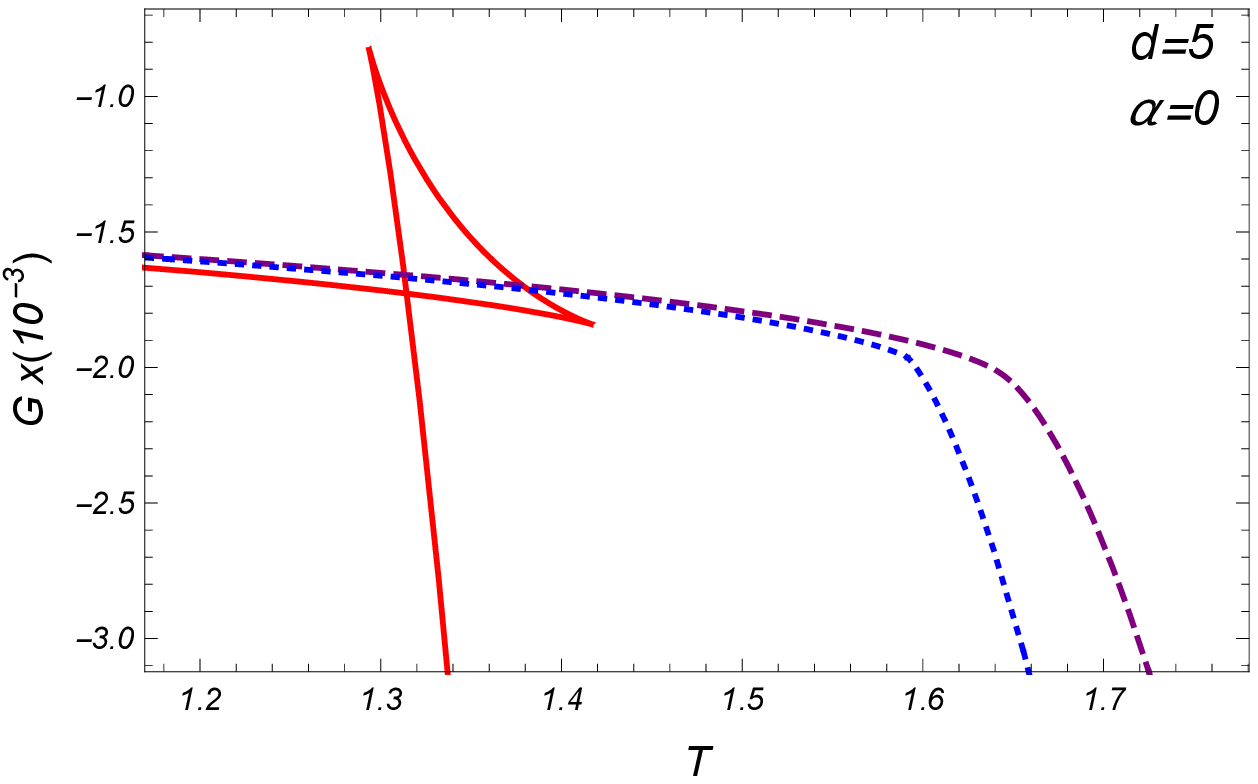}
 		\end{tabbing}
 	\end{center}
 	\caption{Right: The Gibbs free energy as a function of the temperature. Left: Hawking temperature as a function of the horizon radius for Schwarzschild black hole in  $d=5$, with $\alpha=0$.  Here we set $a=0.1$.} 
 	\label{alha0}
 \end{figure}

  Now, we set $\alpha=0$ and solve the equation for critical point by assuming $d>4$. Then, we obtain,
  \begin{align}\label{criticalpointuncharged1}
  r_c = a^{\frac{1}{d-4}}, \qquad P_c=\frac{\left(d-3\right)\, \left(d-4\right)}{16\, \pi \, r_{c}^{2}},	\qquad T_c=\frac{d-4}{2 \, \pi r_{c}}.
  \end{align}
  The values of $r_c$, $P_c$ and $T_c$ are always positive regardless of the value of $a$, such that $a>0$. Consequently, as illustrated by Fig.~\ref{alha0},  the critical behavior for all dimensions greater than $4$ ($d\ge 5$) is recovered if the cloud of strings are introduced. In addition,  the universal number characterizing the equation of state can be read as,
  \begin{align}\label{universalnumber}
 \frac{P_c \, r_c}{T_c}=\frac{d-3}{8},
 \end{align}
as predicted for AdS-Schwarzschild black hole with the above backgrounds  with arbitrary non vanishing string parameter $a$. Adopting the identification proposed for the specific volume $\upsilon$ \cite{Kubiznak}, we derive an interesting relation for $d=6$ : 
\begin{align}\label{universalnumberVdW}
\frac{P_c \, \upsilon_c}{T_c}=\frac{3}{8},
\end{align}
This is the universal number of a Van der Waals fluid,  indicating that $d=6$ black hole in cloud of strings background deserves special attention and analysis.
   
   \subsection{AdS-Schwarzschild   BH  for $d \ge 5$ with  specific choices  of $\omega_q$} 
  With the power of $\omega_q$ present in the state  equation, it is highly non trivial to analytically uncover critical point. However, analytic solutions can well be determined for some particular values  of the  $\omega_q$ parameter, as detailed subsequently.

\begin{figure}[h]
 	\begin{center}
 		\includegraphics[scale=0.62]{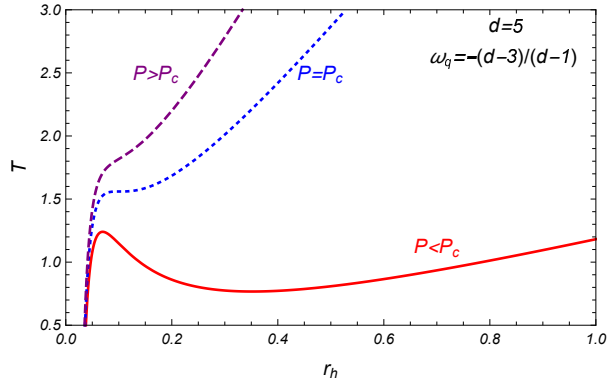}
 	\end{center}
 	\caption{ The  Hawking temperature   as a function of horizon  radius for uncharged black hole in $d=5$ with. We set $\alpha<1$ and $\omega_q=-\left(d-3\right)/\left(d-1\right)$.} 
 	\label{choiceomega2}
 \end{figure}

 For $\omega_q=-\left(d-3\right)/\left(d-1\right)$,  one can show that the critical point is given by:
 \begin{align}\label{criticalpointchoice2}
 r_c = \left(\frac{1-\alpha}{a}\right)^{\frac{1}{4-d}}, \qquad P_c=\frac{\left(d-3\right)\, \left(d-4\right)\, \left(1-\alpha\right)}{16\, \pi \, r_{c}^{2}},	\qquad T_c=\frac{(d-4)\, \left(1-\alpha\right)}{2 \, \pi r_{c}}.
 \end{align} 
 Thus the phase transition is revealed when $\alpha<1$ (see figure~\ref{choiceomega2}).  This condition ensures that the critical point is within the domain where $T_c,\ P_c$ and $r_c$ are strictly positive. Besides,  the universal number kept fixed to its standard value, 
  \begin{align}\label{universalnumber2}
 \frac{P_c \, r_c}{T_c}=\frac{d-3}{8},
 \end{align}
  and hence not affected neither by the presence of quintessence nor by the cloud of strings background, as in the case with $\alpha=0$. This means that  $\omega_q=-\left(d-3\right)/\left(d-1\right)$ restores the predicted universality for $\alpha=0$.

 For $\omega_q=-\left(d-2\right)/\left(d-1\right)$, the solution becomes
\begin{align}\label{criticalpointchoice1}
r_c = a^{\frac{1}{d-4}}, \qquad P_c=\frac{\left(d-3\right)\, \left(d-4\right)}{16\, \pi \, r_{c}^{2}},	\qquad T_c=\frac{d-4}{2 \, \pi r_{c}}-\frac{d-2}{4\, \pi}\alpha.
\end{align}

\begin{figure}[h]
	\begin{center}
		\includegraphics[scale=0.62]{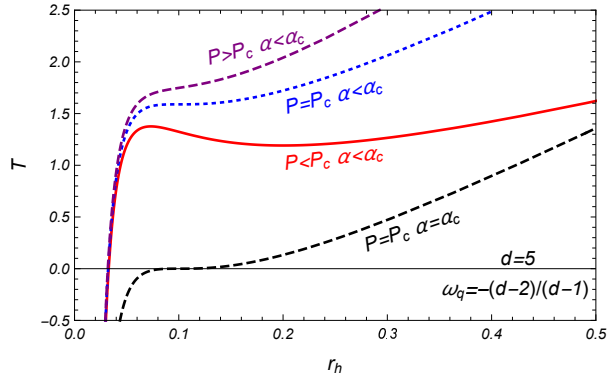}
	\end{center}
	\caption{ The Hawking temperature  as a function of horizon  radius  for AdS-Schwarzschild black hole in $d=5$ for different $\alpha$ with $\omega_q=-\left(d-2\right)/\left(d-1\right)$. } 
	\label{choiceomega}
\end{figure}

In this case, the critical radius and pressure are not affected by the  quintessence presence.  Indeed, the above choice of $\omega_q$ compensates the contribution of $\alpha$ in the state equation and appears as a shifted constant. However, in order to get a Van der Waals phase transition as indicated in Fig.~\ref{choiceomega},  the following additional condition must be fulfilled:
\begin{align}\label{criticalpointchoice1condition}
\alpha=\alpha_c< \frac{2 (d-4)}{d-2} \, a^{\frac{1}{4-d}}.
\end{align}
It is worth to note here that , a similar result of the condition~\eqref{criticalpointchoice1condition}, was  recently obtained for $d=5$ AdS-Schwarzschild black holes in massive gravity theory~\cite{Ghanaatian}.

\begin{figure}[h]
 	\begin{center}
 			 \includegraphics[scale=0.62]{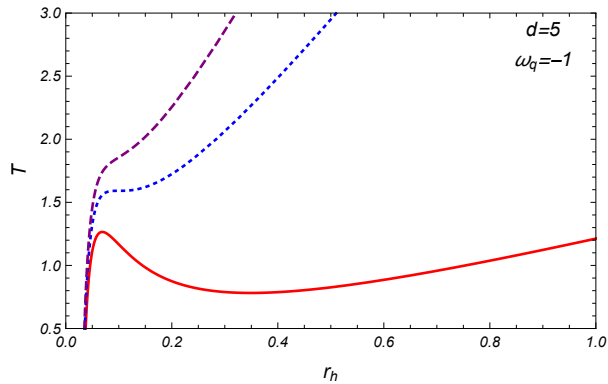}
 	\end{center}
 	\caption{ The  Hawking temperature  as a function of horizon  radius  for AdS-Schwarzschild black hole for $d=5$ and  $\omega_q=-1$. } 
 	\label{choiceomega1}
 \end{figure}

Finally,  assuming $\omega_q=-1$, we get the following solution,  
  \begin{align}\label{criticalpointchoice3}
 r_c = a^{\frac{1}{d-4}}, \qquad P_c=\frac{\left(d-3\right)\, \left(d-4\right)}{16\, \pi \, r_{c}^{2}}+\frac{(d-2)\,(d-1)\,\alpha}{16\pi},	\qquad T_c=\frac{d-4}{2 \, \pi r_{c}}.
 \end{align}
The situation here is opposite to that of $ \omega = - (d-2) / (d-1)$ case. We clearly see that quintessence has no effect on critical radius and temperature, but only affects the pressure. Since $ \alpha $ is positive, we can recover the thermodynamics behavior of a liquid-gas Van der Waals system without any bound or condition related to the values of $ a $ and $ \alpha $. Fig.~\ref{choiceomega1} illustrates the plot of state equation in the $ T-r_h $ plane.

  \FloatBarrier
  \section{Conclusion}\label{sec7}
  In this paper we have studied the charged AdS black hole surrounded by quintessence with cloud of strings present in higher dimensional spacetime. We have generalized the exact solution corresponding of $4d$ spacetime to arbitrary dimension and derived different thermodynamic quantities, such as  the temperature, entropy, Gibbs free energy and heat capacity, which account for quintessence and strings cloud parameters.
  
  In addition, we have studied the critical behavior of these black holes in the extended phase space. As a result, we have observed two main features: \\
  - Phase transition is revealed for all dimensions \\
  - Spacetime dimension modifies the width of coexistence zone.
  \\
  
  We have analyzed the heat capacity to confirm that both small and large black hole phases are stable for all dimensions. However, in the coexistence region, the black hole is locally unstable. The geometrical perspective was also called upon to study the thermodynamic phase transition. We find that the scalar curvature corresponding to Quevedo metric diverges exactly at the singular points signaled by the heat capacity, which means that Quevedo geothermodynamics approach is well appropriate to uncover criticality features of these black holes. Finally, the relevant critical exponents have been calculated and the obtained results are similar to those reported for other black hole solutions in literature. This unexpectedly suggests that the black hole critical behavior in the extended space is insensitive to the quintessence background or the string clouds presence whatever the spacetime dimension.
  \\
  
In the end, we discussed the effects of a cloud of strings and quintessence on the criticality  of uncharged AdS black holes for some specific choices of the quintessence state parameter $\omega_q$.  We found that cloud of strings can induce small/large black hole phase transition for $d\ge 5$, even when $q=0$. Besides, we have shown that if the quintessence is also present, then an additional constraint on  $ a $  parameter must be fulfilled to maintain small/large BH phase transition. As a byproduct, we found that the universal number of the state equation for AdS-Schwarzschild black hole surrounded by a cloud of strings coincides with that of a Van der Waals system ($3/8$) in $d = 6$, as clearly illustrated in Table~\ref{recapetulation}.
\begin{table}
	\centering
	\begin{tabular}{ccccccc}
		\hline\hline 
		$d= 4$ &  $\alpha \ne 0$ & & Hawking-Page phase transition  & $a<1$ \\ 
		\hline
		\multirow{4}{*}{$d\ge 5$ } &  $\alpha =0$ & & SBH/LBH phase transition & For all $a$ and $\alpha$ \\
		\cline{2-5} 
		 &  \multirow{3}{*}{$\alpha \ne 0 $ }  & $ \omega_q=-(d-3)/(d-1)$ & SBH/LBH phase transition & $\alpha<1$ \\ 
		                    &     & $\omega_q=-(d-2)/(d-1)$ & SBH/LBH phase transition & $\alpha< \frac{2 (d-4)}{d-2} \, a^{\frac{1}{4-d}}.$ \\ 
		  				&		& $\omega_q=-1$ & SBH/LBH phase transition & For all $a$ and $\alpha$ \\ 
	
		\hline
		\hline 
	\end{tabular} 
	\caption{Summary of the effects due to dimension, cloud of strings  and quintessence on the thermodynamics phase transition of AdS Schwarzschild black hole} \label{recapetulation}
\end{table}

\end{document}